\documentclass[jcrb]{FBSart}
\usepackage{amsfonts}
\usepackage{amssymb}
\usepackage{amsmath}
\usepackage{graphicx}

\newcommand{\be}{\begin{equation}}
\newcommand{\ee}{\end{equation}}
\renewcommand{\l}{\left}
\renewcommand{\r}{\right}

\newcommand{\chiF}{\chi_F}
\newcommand{\chiG}{\chi_G}
\newcommand{\LQCD}{{\Lambda_{\text{QCD}}}}
\newcommand{\LQCDsq}{{\Lambda_{\text{QCD}}^2}}
\newcommand{\LUVsq}{{\Lambda^2}}
\newcommand{\intq}[1]{\int_{{#1}}}
\newcommand{\gs}[1]{\not\!\text{#1}}

\newcommand{\Trace}{\mbox{\sf Tr}}
\newcommand{\MeV}{\text{MeV}}

\allowdisplaybreaks[4]
\date{March 14, 2003}

\title{
Two-loop Improved Truncation of the Ghost-Gluon Dyson-Schwinger Equations: Multiplicatively Renormalizable Propagators and Nonperturbative Running Coupling
}

\author{J.C.R. Bloch\thanks{\textit{E-mail address}: bloch@wias-berlin.de}}

\institute{DFG Research Center "Mathematics for Key Technologies", c/o  Weierstrass Institute for Applied Analysis and Stochastics, Mohrenstrasse 39, D-10117 Berlin, Germany}

\runningauthor{J.C.R. Bloch}
\runningtitle{
Two-loop Improved Truncation of the Ghost-Gluon DS Equations
}

\begin{document}
\bibliographystyle{h-physrev4}   

\maketitle

\begin{abstract}
The coupled Dyson-Schwinger equations for the gluon and ghost propagators are investigated in the Landau gauge using a two-loop improved truncation that preserves the multiplicative renormalizability of the propagators.
In this truncation all diagrams contribute to the leading order infrared analysis. The infrared contributions of the nonperturbative two-loop diagrams to the gluon vacuum polarization are computed analytically, and this reveals that infrared power behaved propagator solutions only exist when the squint diagram contribution is taken into account. 
For small momenta the gluon and ghost dressing functions behave respectively like $(p^2)^{2\kappa}$ and $(p^2)^{-\kappa}$, and the running coupling exhibits a fixed point.  The values of the infrared exponent and fixed point depend on the precise details of the truncation. The coupled ghost-gluon system is solved numerically for all momenta, and the solutions have infrared behaviors consistent with the predictions of the infrared analysis. 
For truncation parameters chosen such that $\kappa=0.5$, the two-loop improved truncation is able to produce solutions for the propagators and running coupling which are in very good agreement with recent lattice simulations. 
\end{abstract}

\section{Introduction}

In the Standard Model of the strong, weak and electromagnetic forces, the interactions are quantitatively described by gauge field theories. Quantum chromodynamics is a non-Abelian SU(3) gauge theory, and the proof of its renormalizability \cite{'tHooft:1971fh} and discovery of ultraviolet asymptotic freedom \cite{Politzer:1973fx} have been milestones in its acceptance as theory of the strong interaction. For large momenta, perturbation theory seems an appropriate calculational tool, as the coupling becomes small. However, for small momenta the coupling grows large and adequate methods have to be used to study nonperturbative phenomena like confinement, chiral symmetry breaking and fermion mass generation \cite{Alkofer:2000wg}. One such method is the study of the Dyson-Schwinger equations (DSE) \cite{Roberts:1994dr}, and their phenomenological applications to hadronic physics is a subject of growing interest \cite{Roberts:2000aa, Maris:2003vk}. The gluon self-interaction and existence of ghost fields are remarkable features of non-Abelian gauge theories, and the study of the infrared behavior of the gluon and ghost propagators, and of the running coupling has become a major topic in the quest to understand the nonperturbative phenomena in QCD. 

Early studies of the Dyson-Schwinger equation for the gluon propagator in the Landau gauge concluded that the gluon propagator is highly singular in the infrared \cite{Mandelstam:1979xd, Atkinson:1981er, Atkinson:1982ah, Brown:1989bn}. However, these studies neglected any contribution of the ghost fields, and posited cancellations of certain leading terms in the equations. It is therefore far from certain that these solutions possess the correct infrared behavior of QCD. Nevertheless, these solutions are at the origin of successful phenomenological applications, which can be understood by observing that they generate the necessary integration strength in the kernels of the gap and Bethe-Salpeter equations \cite{Hawes:1998cw}.

The important role played by the coupling of ghost and gluon fields to generate a consistent infrared behavior of QCD in the Landau gauge was first demonstrated in a study of the \textit{coupled} propagator equations \cite{vonSmekal:1998is}. The propagator dressing functions were found to behave like powers for small momenta: the gluon as $(p^2)^{2\kappa}$ and the ghost as $(p^2)^{-\kappa}$. Furthermore, the strong running coupling is related to both propagators and exhibits an infrared fixed point $\alpha_0$ whose value depends on $\kappa$. 
Initial studies \cite{vonSmekal:1998is, Atkinson:1998tu} truncated the system using \textit{Ans\"atze} for the ghost-gluon and three-gluon vertices, neglecting the two-loop gluon vacuum polarization diagrams and introducing an angular averaging procedure. The infrared analysis was performed analytically and yielded $\kappa=0.92$, $\alpha_0=9.5$ \cite{vonSmekal:1998is}, and $\kappa=0.77$, $\alpha_0=11.5$ \cite{Atkinson:1998tu}, depending on the details of the truncation. The coupled equations were also solved numerically for all momenta, and the numerical solutions have the infrared behavior predicted by the analytic infrared analysis. The uniqueness of the solutions of ref. \cite{Atkinson:1998tu} was recently confirmed through an extended analysis of the infrared and ultraviolet asymptotic behaviors of the truncation \cite{Kondo:2002cx, Kondo:2002cy}. A discussion of the Faddeev-Popov quantization at the nonperturbative level and of the Gribov horizon problem can be found in ref.\ \cite{Zwanziger:2001kw}. Comparison with lattice simulations \cite{Bonnet:2000kw, Langfeld:2001cz, Bloch:2002we} is quite encouraging as these studies have revealed quite similar infrared behaviors for the propagators, i.e. a vanishing gluon dressing function and a singular ghost dressing function. Nevertheless, quantitative discrepancies in the value of $\kappa$, fixed point of the coupling and behavior of the gluon propagator at intermediate momentum hint to the need of improved truncations in the Dyson-Schwinger equation studies.

As was shown in ref.\ \cite{Atkinson:1998zc} the recipe used for angular averaging influences the value of the exponent $\kappa$ and even the existence of infrared power law solutions, and a method was developed therein to treat the angular integrals exactly. The infrared analysis revealed an exponent $\kappa=1$, agreeing with the Gribov form of the gluon and ghost propagators \cite{Gribov:1978wm}, and a smaller fixed point value $\alpha_0=4.2$. Unfortunately, the numerical analysis could not find solutions satisfying the integral equations over the whole momentum range, and the question whether this infrared behavior is indeed a property of QCD or a spurious solution of the infrared analysis was left unanswered. Clearly, the existence of power laws satisfying the infrared consistency conditions is not sufficient to ensure the existence of an overall solution of the integral equations. 

To avoid spurious unrenormalizable ultraviolet quadratic divergences, refs.\ \cite{vonSmekal:1998is, Atkinson:1998tu, Atkinson:1998zc} extracted the gluon dressing function by contracting the gluon equation with the Brown-Pennington (BP) tensor \cite{Brown:1989bn}. In more recent work \cite{Lerche:2001, Lerche:2002ep} the contraction was performed using an arbitrary tensor, and the infrared analysis yielded an additional solution for $\kappa$ slightly larger than 0.5  when using a \textit{non}-BP tensor. These studies were restricted to an infrared analysis, and no full numerical solution was given. In fact, a straightforward analysis shows that these truncations are plagued by unrenormalizable ultraviolet quadratic divergences, and, as will be shown later in this paper, the new infrared solution for $\kappa$ is spurious as its existence is directly related to that of the ultraviolet quadratic divergences.

This truncation was later enhanced \cite{Fischer:2002hn} by removing the ultraviolet quadratic divergences using a prescription specially constructed to leave the infrared analysis of ref.\ \cite{Lerche:2001} unchanged, thus retaining the spurious $\kappa \gtrapprox 0.5$ solution. The merit of this study is to produce the first numerical solution to the integral equations with full angular integrals.  However, as will be discussed later, its prescription to remove the quadratic divergences is not satisfying, and subtracting these in a consistent way removes the spurious infrared solution and destroys the existence of the numerical solution.

For all the above-mentioned truncations, following in the footsteps of ref.\ \cite{vonSmekal:1998is}, only the ghost loop contributes to the leading infrared power of the gluon DSE. At the same time, all these truncations violate the multiplicative renormalizability of the propagators.
In ref.\ \cite{Bloch:2001wz} I proposed a novel truncation of the Dyson-Schwinger equations that respects the multiplicative renormalizability of the propagators and satisfies their resummed leading order perturbative behavior. A major new feature of the truncation is that all vacuum polarization diagrams of the gluon DSE contribute to its leading infrared power.
An analytic infrared analysis of the coupled ghost-gluon system was performed taking into account the ghost and gluon loops in the gluon vacuum polarization, but still neglecting the two-loop diagrams. The study could not find infrared power laws simultaneously solving the gluon and ghost equations.  
Although the two-loop diagrams also contribute to the leading infrared power, their numerical contribution could not yet be evaluated at that time, and it was suggested that a more detailed study of the missing two-loop contributions is needed to get a better understanding of the infrared behavior of QCD. 

In the present paper the MR truncation of ref.\ \cite{Bloch:2001wz} for the coupled ghost-gluon DSEs in pure Yang-Mills theory is generalized further. The various infrared contributions to the ghost self-energy and gluon vacuum polarization are computed analytically. This includes an analytic solution for the two-loop infrared contributions of the sunset and squint diagrams. The contribution of the squint diagram proves to be such that propagator power laws satisfy the infrared consistency conditions imposed by ghost and gluon equations, and the values of the infrared exponent and fixed point of the coupling depend on the details of the truncation.

For the numerical study, the two-loop integrals are replaced by effective one-loop integrals with same infrared behavior and consistent subleading ultraviolet behavior. The equations are then solved numerically for {\it all} momenta, and the numerical solutions agree with the predicted infrared and ultraviolet behaviors. The results for ghost and gluon propagators, and for the running coupling are also compared to recent results from lattice simulations \cite{Bloch:2002we}.

\section{The Equations}
\label{Sect:Equations}

The Dyson-Schwinger equations for the ghost and gluon propagators can be diagrammatically written as
\begin{align}
\label{diagghost}\begin{split}&\includegraphics[width=9cm]{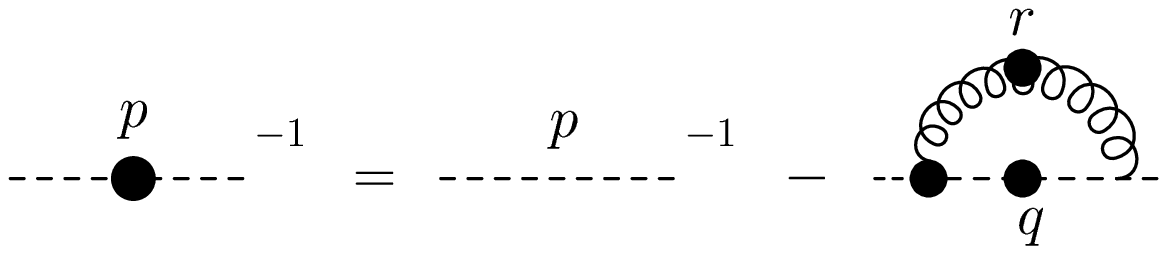}\end{split} \\
\label{diaggluon}\begin{split}&\includegraphics[width=12cm]{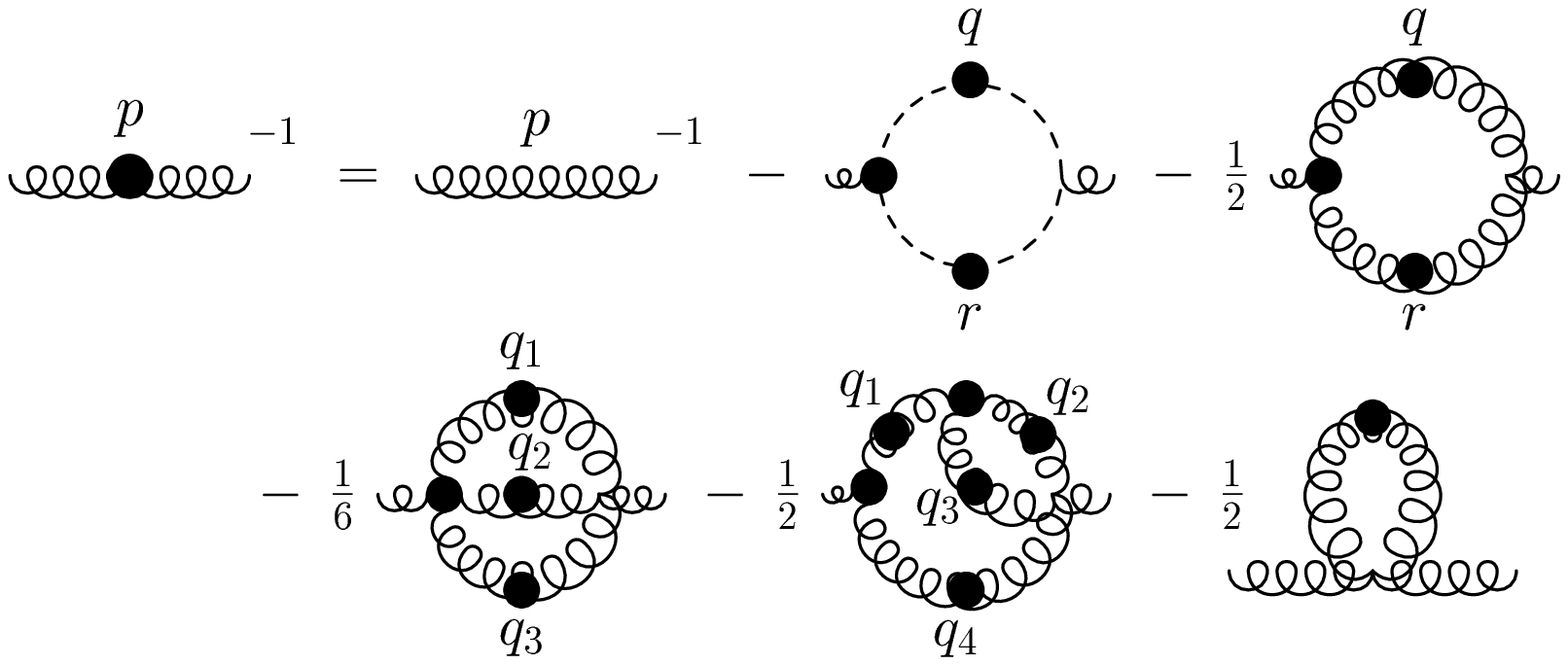} \end{split}
\end{align}
with the labelled momenta flowing from left to right. Several contributions to the gluon vacuum polarization were neglected in previous studies, and here the scope of the investigation is extended by considering its complete form including the contributions from the ghost loop, gluon loop, three-gluon loop and four-gluon loop\footnote{The tadpole contribution is independent of the external momentum, and vanishes after subtraction of the renormalized equations.}.  Written out explicitly, the Euclidean space equations are
\begin{align}
\label{SD-Gh-full}
\l[\Delta^{ab}(p)\r]^{-1} &= \l[\Delta^{0,ab}(p)\r]^{-1} - \Sigma^{ab}(p) \,,\\
\label{SD-Gl-full}
\l[D_{\mu\nu}^{ab}(p)\r]^{-1} &= \l[D^{0,ab}_{\mu\nu}(p)\r]^{-1} - \Pi_{\mu\nu}^{ab}(p) \,,
\end{align}
with ghost self-energy integral\footnote{$\intq{q} = \int d^4 q/(2\pi)^4$}
\be
\Sigma^{ab}(p) = \intq{q}  \, G^{a_2 a a_1}_\mu(p,q) \Delta^{a_1 b_1}(q) G_\nu^{0, b_2 b_1 b}(q,p) D_{\mu\nu}^{b_2 a_2}(r) \,,
\ee
and gluon vacuum polarization
\be
\Pi_{\mu\nu}^{ab}(p) = \Pi_{\mu\nu}^{2c,ab}(p) + \Pi_{\mu\nu}^{2g,ab}(p) + \Pi_{\mu\nu}^{3g,ab}(p) + \Pi_{\mu\nu}^{4g,ab}(p) \,,
\ee
where
\begin{align}
\label{pight}
\Pi_{\mu\nu}^{2c,ab}(p) &=
(-1) \intq{q} \, G^{a a_2 a_1}_\mu(-r, q) \Delta^{a_1 b_1}(q) G^{0, b b_1 b_2}_\nu(q, -r) \Delta^{b_2 a_2}(-r) \,,\\
\label{piglt}
\Pi_{\mu\nu}^{2g,ab}(p) &=
\frac{1}{2} \intq{q} \, \Gamma^{3g,a a_2 a_1}_{\mu\alpha_2\alpha_1}(p,-r,-q) D_{\alpha_1\beta_1}^{a_1 b_1}(q) \Gamma^{3g,0,b b_1 b_2}_{\nu\beta_1\beta_2}(-p,q,r) D_{\beta_2\alpha_2}^{b_2 a_2}(r) \,,\\
\label{pi3glt}
\begin{split}
\Pi^{3g,ab}_{\mu\nu}(p) &=  \frac{1}{6} \intq{q_1}\intq{q_2} \, \Gamma^{4g,a a_3 a_2 a_1}_{\mu\alpha_3\alpha_2\alpha_1}(p,-q_3,-q_2,-q_1) D_{\alpha_1\beta_1}^{a_1 b_1}(q_1) D_{\alpha_2\beta_2}^{a_2 b_2}(q_2) \\ 
& \quad\times D_{\alpha_3\beta_3}^{a_3 b_3}(q_3) \Gamma^{4g,0,b b_1 b_2 b_3}_{\nu\beta_1\beta_2\beta_3}(-p,q_1,q_2,q_3) \,,
\end{split}\\
\label{pi4glt}
\begin{split}
\Pi^{4g,ab}_{\mu\nu}(p) &=  \frac{1}{2} \intq{q_1}\intq{q_2} \Gamma^{3g, a a_4 a_1}_{\mu\alpha_4\alpha_1}(p,-q_4,-q_1) 
D_{\alpha_1\beta_1}^{a_1 b_1}(q_1) \Gamma^{3g, b_1 a_3 a_2}_{\beta_1\alpha_3\alpha_2}(q_1,-q_3,-q_2)
\\
& \quad \times  D_{\alpha_2\beta_2}^{a_2 b_2}(q_2) D_{\alpha_3\beta_3}^{a_3 b_3}(q_3) 
 \Gamma^{4g,0,b b_2 b_3 b_4}_{\nu\beta_2\beta_3\beta_4}(-p,q_2,q_3,q_4) D_{\beta_4\alpha_4}^{b_4 a_4}(-q_4) \,.
\end{split}\raisetag{16pt}
\end{align}
Here $D$ is the gluon propagator, $\Delta$ the ghost propagator, $\Gamma^{3g}$ the three-gluon vertex, $\Gamma^{4g}$ the four-gluon vertex, $G$ the ghost-gluon vertex, the zero superscript denotes bare quantities, and the definitions of the momenta can be read off from the diagrammatic Eqs. (\ref{diagghost}) and (\ref{diaggluon}). The extra factor $(-1)$ in front of the ghost loop (\ref{pight}) is due to the Fermi statistics of the ghost field. The momenta in the three- and four-gluon vertices are incoming momenta, the ghost momentum in the ghost-gluon vertex is incoming, and the anti-ghost momentum is outgoing. The integrals are regulated using an ultraviolet cutoff $\Lambda$.

The general expressions for the full gluon and ghost propagators in a covariant gauge $\xi$ can be written as
\begin{align}\label{Full-Gl}
D_{\mu\nu}^{ab}(p) &=  \left[ \left(g_{\mu\nu}-\frac{p_\mu p_\nu}{p^2}\right) \frac{F(p^2)}{p^2} + \xi \frac{p_\mu p_\nu}{p^4} \right] \delta^{ab} \,,\\
\label{Full-Gh}
\Delta^{ab}(p) &= -\frac{G(p^2)}{p^2} \,\delta^{ab}\,,
\end{align}
where the scalar functions $F(p^2)$ and $G(p^2)$ are the gluon and ghost dressing functions, and the bare propagators correspond to $F(p^2)=1$ and $G(p^2)=1$. The full vertices can be decomposed in their color structures:
\begin{align}
\label{cgV}
G^{abc}_\mu(p,p') &= i g_0 f^{abc} p'_\lambda G_{\lambda\mu}(p,p') \,, \\
\label{3gV}
\Gamma^{3g,abc}_{\mu\nu\rho}(p,q,r) &= - i g_0 f^{abc} \Gamma^{3g}_{\mu\nu\rho}(p,q,r) \,, \\
\label{4gV}
\begin{split}
\Gamma^{4g,abcd}_{\mu\nu\rho\sigma}(p,q,r,s) &= - g_0^2 [ f^{abe}f^{cde}\Gamma^{4g}_{\mu\nu\rho\sigma}(p,q,r,s) + f^{ace}f^{bde} \Gamma^{4g}_{\mu\rho\nu\sigma}(p,r,q,s) \\ 
& \qquad + f^{ade}f^{bce}\Gamma^{4g}_{\mu\sigma\nu\rho}(p,s,q,r) ] \,,
\end{split}
\end{align} 
where $g_0$ is the bare coupling and $p'$ is the anti-ghost momentum flowing out of the ghost-gluon vertex. After separation of the color structure, the tensor $\Gamma^{3g}$ is antisymmetric under exchange of any two arguments, while $\Gamma^{4g}$ is antisymmetric under exchange of $1 \leftrightarrow 2$ or $3 \leftrightarrow 4$, and symmetric under exchange of $(1,2) \leftrightarrow (3,4)$.
The bare vertices correspond to
\begin{align}
\label{G}
G_{\lambda\mu}^{0}(p,p') &= \delta_{\lambda\mu} \,, \\
\label{Gam3}
\Gamma^{3g, 0}_{\mu \nu \rho}(p,q,r) 
	&= ({p}-{q})_{\rho} \delta_{\mu \nu} +({q}-{r})_{\mu} \delta_{\nu \rho} + ({r}-{p})_{\nu} \delta_{\rho \mu} \,,\\
\label{Gam4}
\Gamma^{4g, 0}_{\mu \nu \rho \sigma} (p,q,r,s) &= 
\delta_{\mu \rho} \delta_{\nu \sigma} - \delta_{\mu \sigma} \delta_{\nu \rho} \,,
\end{align}
in Eqs.\ \eqref{cgV}, \eqref{3gV} and \eqref{4gV}.

The Greens functions are potentially divergent as the ultraviolet cutoff is taken to infinity, and therefore renormalized fields and couplings are introduced. 
Proceeding as described in ref.\ \cite{Bloch:2001wz}, the exact equations for the renormalized dressing functions in the Landau gauge can be written as
\begin{align}
\label{SD-Gh}
\frac{1}{G_R(p^2,\mu^2)} &= \tilde Z_3(\mu^2,\Lambda^2) - \Sigma(p^2) \,,\\
\label{SD-Gl}
\frac{1}{F_R(p^2,\mu^2)} &= Z_3(\mu^2,\Lambda^2) - \Pi(p^2)  \,,
\end{align}
where the renormalized and unrenormalized dressing functions are related through multiplicative renormalizability:
\be\label{Z3}
Z_3(\mu^2,\Lambda^2) = \frac{F(p^2,\Lambda^2)}{F_R(p^2,\mu^2)} \quad , \qquad
\tilde Z_3(\mu^2,\Lambda^2) = \frac{G(p^2,\Lambda^2)}{G_R(p^2,\mu^2)} \quad,
\ee
and $Z_3$ and $\tilde Z_3$ are the gluon and ghost field renormalization constants fixed by the renormalization conditions $F_R(\mu^2,\mu^2)$ = $G_R(\mu^2,\mu^2)$ = 1.
The scalar gluon equation \eqref{SD-Gl} was extracted from the tensor equation (\ref{SD-Gl-full}) by contracting with the tensor 
\be\label{project}
{\cal P}_{\mu\nu}(p) = \delta_{\mu\nu}-n \frac{p_\mu p_\nu}{p^2} \,.
\ee
Throughout this paper the Brown-Pennington tensor ($n=4$) will be used, unless stated otherwise, such that spurious, unrenormalizable quadratic ultraviolet divergences are avoided \cite{Brown:1989bn}. The ghost self-energy contribution in Eq.\ \eqref{SD-Gh} is given by
\be
\label{SE-Ghb}
\Sigma(p^2) = \frac{N_c \, g^2_\mu}{Z_3 \tilde Z_3} \intq{q}\, T(p^2,q^2,r^2)\, G(q^2) F(r^2) \,,
\ee
and the gluon vacuum polarization in Eq.\ \eqref{SD-Gl} by
\be\label{vacpoltot}
\Pi(p^2) = \Pi^{2c}(p^2) + \Pi^{2g}(p^2) + \Pi^{3g}(p^2) + \Pi^{4g}(p^2) \,,
\ee
with
\begin{align}
\label{pighb}
\Pi^{2c}(p^2) &= \frac{N_c \, g^2_\mu}{\tilde Z_3^2} \intq{q} \,  M(p^2,q^2,r^2) \,  G(q^2) G(r^2)\,,\\
\label{piglb}
\Pi^{2g}(p^2) &= \frac{N_c \, g^2_\mu}{\tilde Z_3^2} \intq{q} \, Q(p^2,q^2,r^2) \,  F(q^2) F(r^2)\,,\\
\label{3gloopb}
\Pi^{3g}(p^2) &= \frac{N_c^2 g^4_\mu}{Z_3 \tilde Z_3^4} \intq{q_1} \intq{q_2}  \, Q_3(p,q_1,q_2,q_3) \,  F(q_1^2) F(q_2^2) F(q_3^2)\,,\\
\label{4gloopb}
\Pi^{4g}(p^2) &= \frac{N_c^2 g^4_\mu}{Z_3 \tilde Z_3^4} \intq{q_1} \intq{q_2} \,  Q_4(p,q_1,q_2,q_3,q_4)\, F(q_1^2)  F(q_2^2) F(q_3^2) F(q_4^2) \,,
\end{align}
where the contractions over color indices were performed using the formulae of \ref{App:colorcontractions}, and $N_c$ is the number of colors. The renormalized coupling $g_\mu$ was introduced through the renormalization of the ghost-gluon interaction
\be
g_\mu = g(\mu^2) = \frac{Z_3^{1/2}(\mu^2,\Lambda^2)\tilde
Z_3(\mu^2,\Lambda^2)}{\tilde{Z}_1(\mu^2,\Lambda^2)} \, g_0(\Lambda^2) \,,
\label{renormg}
\ee
where $\tilde Z_1$ is the ghost-gluon vertex renormalization constant. Equations \eqref{SE-Ghb}-\eqref{4gloopb} make use of the universality of the strong coupling stating that the renormalized coupling is identical for the ghost-gluon, three-gluon and four-gluon interaction. Furthermore, $\tilde Z_1 \equiv 1$ in the Landau gauge, as shown by Taylor \cite{Taylor:1971ff}. 
In the Landau gauge the various kernels in Eqs.\ \eqref{SE-Ghb}-\eqref{4gloopb} are defined by
\begin{align}
\label{kernelT} T(p^2,q^2,r^2) &= - \frac{1}{p^2 \, q^2 \, r^2} \, g^\perp_{\mu\nu}(r) \, G_\mu(p,q) \, G^0_\nu(q,p) \,,\\
\label{kernelM} M(p^2,q^2,r^2) &=  \frac{1}{3 \, p^2 \,q^2\, r^2} \,  {\cal P}_{\mu\nu}(p) \,G_\mu(-r, q) \, G^0_\nu(q, -r) \,,\\
\begin{split}
\label{kernelQ} Q(p^2,q^2,r^2) &= - \frac{1}{6 \, p^2 q^2 r^2} \, {\cal P}_{\mu\nu}(p) \,   g^\perp_{\alpha_1\beta_1}(q) \, g^\perp_{\alpha_2\beta_2}(r) \\ 
& \times  \Gamma^{3g}_{\mu\alpha_2\alpha_1}(p,-r,-q) \, \Gamma^{3g,0}_{\nu\beta_1\beta_2}(-p,q,r) \,,
\end{split}\\
\label{kernelQ3}
\begin{split}
Q_3(p,q_1,q_2,q_3) &=  \frac{1}{36 \, p^2 q_1^2 q_2^2 q_3^2} {\cal P}_{\mu\nu}(p) \, 
        g^\perp_{\alpha_1\beta_1}(q_1) g^\perp_{\alpha_2\beta_2}(q_2) g^\perp_{\alpha_3\beta_3}(q_3) \\
& \times [
       ( 2 \Gamma^{4g}_{\mu\alpha_1\alpha_2\alpha_3}(p,-q_1,-q_2,-q_3) \Gamma^{4g,0}_{\nu\beta_1\beta_2\beta_3}(-p,q_1,q_2,q_3) \\
	& - \Gamma^{4g}_{\mu\alpha_1\alpha_2\alpha_3}(p,-q_1,-q_2,-q_3) \Gamma^{4g,0}_{\nu\beta_2\beta_3\beta_1}(-p,q_2,q_3,q_1) \\
	& - \Gamma^{4g}_{\mu\alpha_1\alpha_2\alpha_3}(p,-q_1,-q_2,-q_3) \Gamma^{4g,0}_{\nu\beta_3\beta_1\beta_2}(-p,q_3,q_1,q_2) )  \\
	& + \textrm{cyclic perm. } (1,2,3) ] \,,
\end{split} \\
\label{kernelQ4}
\begin{split}
Q_4(p,q_1,q_2,q_3,q_4) &= \frac{1}{12 \, p^2 q_1^2 q_2^2 q_3^2 q_4^2} \, {\cal P}_{\mu\nu}(p)  g^\perp_{\alpha_1\beta_1}(q_1) g^\perp_{\alpha_2\beta_2}(q_2)  g^\perp_{\alpha_3\beta_3}(q_3)  \\ 
& \times g^\perp_{\beta_4\alpha_4}(-q_4) \Gamma^{3g}_{\mu\alpha_4\alpha_1}(p,-q_4,-q_1) \Gamma^{3g}_{\beta_1\alpha_3\alpha_2}(q_1,-q_3,-q_2) \\ 
& \times [ \Gamma^{4g,0}_{\nu\beta_2\beta_3\beta_4}(-p,q_2,q_3,q_4) + \Gamma^{4g,0}_{\nu\beta_3\beta_4\beta_2}(-p,q_3,q_4,q_2) \\
& \quad - 2 \, \Gamma^{4g,0}_{\nu\beta_4\beta_2\beta_3}(-p,q_4,q_2,q_3) ] \,,
\end{split}
\end{align}
where $g_{\mu\nu}^\perp(q) = g_{\mu\nu} - q_\mu q_\nu/q^2$. The kernels \eqref{kernelT} to \eqref{kernelQ4} contain full vertices which depend on the propagator functions $F$ and $G$ by the intermediary of the three-point Dyson-Schwinger equations. 

As can be seen from Eqs.\ (\ref{SE-Ghb}) to (\ref{4gloopb}), the renormalization point $\mu$ only enters the right hand sides of Eqs.\ \eqref{SD-Gh} and \eqref{SD-Gl} through the renormalized coupling $g_\mu$ and the renormalization constants $Z_3$ and $\tilde Z_3$. All vertices and dressing functions inside the integration kernels were left unrenormalized on purpose. The reason for this will become clear when the MR truncation is constructed in the next section.

The knowledge of the propagator dressing functions is phenomenologically important because of its relation to the renormalization group invariant running coupling. From the definitions \eqref{Z3}, \eqref{renormg} of the renormalized quantities it is straightforward to show that in the Landau gauge the product
\be
\label{rgi2}
\hat\alpha(q^2,\mu^2) \equiv  \alpha(\mu^2) \, F_R(q^2,\mu^2) \, G_R^2(q^2,\mu^2)
\ee
is renormalization group invariant, i.e.\ independent of $\mu^2$ \cite{Mandelstam:1979xd, vonSmekal:1998is, Atkinson:1998tu}. Hence, the product $\hat\alpha(q^2,\mu^2)$ is nothing else but the nonperturbative running coupling $\alpha(q^2)$, as is easily demonstrated by choosing $\mu^2=q^2$ and noting that $F_R(q^2,q^2) = G_R(q^2,q^2) = 1$. The running coupling and its full nonperturbative behavior will be a major focus in the remaining of this paper.

\section{The Truncation}
\label{Sect:Truncation}

As proposed in ref.\ \cite{Bloch:2001wz} the renormalization constants $Z_3$ and $\tilde Z_3$ in Eqs.\ (\ref{SE-Ghb}) to (\ref{4gloopb}) are eliminated using their definitions \eqref{Z3}, where the ratios of unrenormalized to renormalized dressing functions are independent of the chosen momentum $p$ for a multiplicatively renormalizable theory. This arbitrariness in the choice of $p$ will be used to construct a truncation which not only satisfies the principles of multiplicative renormalizability, but also satisfies the resummed leading order perturbative results. 

After elimination of $Z_3$ and $\tilde Z_3$ using Eq.\ (\ref{Z3}) the full quantum corrections (\ref{SE-Ghb}), \eqref{pighb}-(\ref{4gloopb}) can be rewritten as 
\begin{align}\label{SE-Gh}
\Sigma(p^2)
&= N_c \, g_\mu^2 \intq{q} \, \l[ \frac{G(q^2)F(r^2)}{G^1[\mathbf a_\Sigma]F^1[\mathbf b_\Sigma]} \, T(p^2,q^2,r^2) \r] \, G_R^1[\mathbf a_\Sigma] F_R^1[\mathbf b_\Sigma] \,,\\
\label{pigh}
\Pi^{2c}(p^2) &= N_c \, g_\mu^2  \intq{q} \, \l[ \frac{G(q^2)G(r^2)}{G^2[\mathbf a_{2c}]F^0[\mathbf b_{2c}]} \, M(p^2,q^2,r^2) \r] \,  G_R^2[\mathbf a_{2c}]F_R^0[\mathbf b_{2c}] \,,\\
\label{pigl}
\Pi^{2g}(p^2) &= N_c \, g_\mu^2  \intq{q} \, \l[\frac{F(q^2)F(r^2)}{G^2[\mathbf a_{2g}]F^0[\mathbf b_{2g}]} \, Q(p^2,q^2,r^2)\r]  \,  G_R^2[\mathbf a_{2g}]F_R^0[\mathbf b_{2g}] \,,\\
\label{3gloop}
\begin{split}
\Pi^{3g}(p^2) &= N_c^2 g_\mu^4 \intq{q_1} \intq{q_2} \l[ \frac{F(q_1^2)F(q_2^2) F(q_3^2)}{G^4[\mathbf a_{3g}]F^1[\mathbf b_{3g}]} \, Q_3(p,q_1,q_2,q_3) \r] G_R^4[\mathbf a_{3g}] F_R^1[\mathbf b_{3g}]  \,,
\end{split}\\
\label{4gloop}
\begin{split}
\Pi^{4g}(p^2) &= N_c^2 g_\mu^4 \intq{q_1} \intq{q_2} \l[\frac{ F(q_1^2) F(q_2^2) F(q_3^2) F(q_4^2)}{ G^4[\mathbf a_{4g}]F^1[\mathbf b_{4g}]} \, Q_4(p,q_1,q_2,q_3,q_4) \r] \\ 
& \hspace{7.5cm} \times G_R^4[\mathbf a_{4g}]F_R^1[\mathbf b_{4g}]  \,,
\end{split}
\end{align}
where the symbolic notation
\be\label{Xm}
X^m[\mathbf v] = \prod_{i=1}^{n} X^{v_i}(q_i^2) \quad , \qquad  \sum_{i=1}^{n} v_i=m \,, 
\ee
was introduced, which represents a product of powers of dressing functions evaluated at the internal momenta. Here $X$ stands for any of the dressing functions (renormalized or unrenormalized), $n$ is the number of internal legs in the diagram, $q_i$ enumerates the momenta of the internal legs,  $\mathbf v=(v_1 \ldots v_n)$ is a vector of exponents, and the superscript $m$ gives the total sum of the exponents. For example, the notation $G^4[\mathbf a_{3g}]$ in Eq.\ \eqref{3gloop} can be written out explicitly as
\[
G^4[\mathbf a_{3g}] = G^{a_{3g,1}}(q_1^2) \; G^{a_{3g,2}}(q_2^2) \; G^{a_{3g,3}}(q_3^2) \,,
\]
where the exponents add up as $a_{3g,1} + a_{3g,2} + a_{3g,3} = 4$.  Note that in the single loops \eqref{SE-Gh} to \eqref{pigl} the internal momenta $q_1$ and $q_2$ were respectively labelled as $q$ and $r$.

Equations (\ref{SE-Gh}) to (\ref{4gloop}) are still \textit{exact}, and have the peculiarity that the renormalization point only enters through the renormalized coupling and the renormalized dressing functions, as all the renormalization constants have been eliminated. Furthermore, the objects between square brackets contain regularized, but unrenormalized, quantities which depend on the ultraviolet regulator $\Lambda$, the external momentum and the loop momenta, but are independent of the renormalization point $\mu$.

As was discussed in detail in ref.\ \cite{Bloch:2001wz} the shapes of Eqs.\ (\ref{SE-Gh}) to (\ref{4gloop})  are such that a truncation respecting the multiplicative renormalizability of the propagators and satisfying the leading order resummed perturbative results can be constructed in a straightforward way. The truncation consists in replacing each expression between square brackets in Eqs.\ (\ref{SE-Gh}) to (\ref{4gloop}) by its tree level analog
\begin{align}\label{SE-Ghtrunc}
\Sigma(p^2)
&= N_c \, g_\mu^2 \intq{q}\, T_0(p^2,q^2,r^2) \, G_R^1[\mathbf a_\Sigma] F_R^1[\mathbf b_\Sigma] \,, \\
\label{pightrunc}
\Pi^{2c}(p^2) &= N_c \, g_\mu^2 \intq{q} \, M_0(p^2,q^2,r^2) \,  G_R^2[\mathbf a_{2c}] F_R^0[\mathbf b_{2c}] \,,\\
\label{pigltrunc}
\Pi^{2g}(p^2) &= N_c \, g_\mu^2 \intq{q} \, Q_0(p^2,q^2,r^2) \,  G_R^2[\mathbf a_{2g}] F_R^0[\mathbf b_{2g}]\,,\\
\label{3glooptrunc}
\Pi^{3g}(p^2) &= N_c^2 g_\mu^4 \intq{q_1} \intq{q_2} \,  Q_3^0(p,q_1,q_2,q_3) \, G_R^4[\mathbf a_{3g}] F_R^1[\mathbf b_{3g}] \,,\\
\label{4glooptrunc}
\Pi^{4g}(p^2) &= N_c^2 g_\mu^4 \intq{q_1} \intq{q_2} \,  Q_4^0(p,q_1,q_2,q_3,q_4)\,   G_R^4[\mathbf a_{4g}] F_R^1[\mathbf b_{4g}] \,.
\end{align}
Because of the introduction of the product form \eqref{Xm} the truncation is a somewhat generalized version of that proposed in ref.\ \cite{Bloch:2001wz}, where specific sets of values for the $\mathbf a$'s and $\mathbf b$'s of Eqs.\ \eqref{SE-Ghtrunc}-\eqref{4glooptrunc} were chosen. Note that choice of the sets $\mathbf a$ and $\mathbf b$ can be done independently in the different self-energy and vacuum polarization diagrams, as is suggested by their different subscripts. The proof that the truncation respects multiplicative renormalizability and the leading order resummed perturbative results can be found in ref.\ \cite{Bloch:2001wz}. To ensure that the leading order perturbative results are reproduced it is crucial to choose internal loop momenta when using Eq.\ \eqref{Z3} to eliminate the renormalization constants. This was provided for in the definition \eqref{Xm} of $X^m[\mathbf{v}]$. Choosing a fixed momentum, e.g. the renormalization point $\mu$, would fail to reproduce the perturbative results.

The freedom of choice of the sets $\mathbf a$ and $\mathbf b$ means that even this simple truncation procedure is far from unique. For most diagrams, concerns of naturalness and symmetry will fix the choice in the further treatment. For other diagrams, the sets will be used as parameters in the study of the solutions of the coupled DSEs.
 
Note that the approach developed herein is quite different from that of refs.\ \cite{vonSmekal:1998is, Lerche:2002ep} where various explicit corrections to the ghost-gluon and triple-gluon vertices were studied. Herein I consider effective kernel corrections to the equations, which guarantee that the kernels satisfy properties of MR and PT. 
 
The tree-level kernels of the one-loop diagrams in Eqs.\ \eqref{SE-Ghtrunc}-\eqref{pigltrunc} are
\begin{align}
\label{Txyz0}
T_0(x,y,z) &=
- \left({\frac{{x}}{y}} -
      2 + {\frac{{y}}{x}}\right){\frac{1}{4\,z^2}} +
  \left({\frac{1}{y}} + {\frac{1}{x}}\right){\frac{1}{2\,z}} - {\frac{1}{4\,x\,y}} \,,\\
\label{Mxyz0}
M_0(x,y,z) &= \frac{1}{3}\left[ - \frac{1}{2\,y\,z} - \frac{1 }{2\,x}\left( \frac{1}{y} + \frac{1}{z} \right) + \frac{1}{x^2}\left(\frac{y}{z} - 2 + \frac{z}{y}\right)\right] \,,\\
\label{Qxyz0}
\begin{split}
Q_0(x,y,z) &= \frac{1}{3}\left[ - \frac{x^2}{8\,y^2\,z^2} - x\,\left( \frac{1}{y\,z^2} + \frac{1}{y^2\,z} \right)  + \frac{15}{8\,z^2} +
    \frac{19}{4\,y\,z}  + \frac{15}{8\,y^2} \right. \\ 
     &  \left. - \frac{1}{4 x} \left( \frac{y}{z^2}  - \frac{13}{z} - \frac{13}{y}  + \frac{z}{y^2} \right) 
     - \frac{1}{x^2}\left(\frac{y^2}{2\,z^2} + \frac{4\,y}{z} -9 + \frac{4\,z}{y} + \frac{z^2}{2\,y^2} \right) \right] \,,
\end{split}\raisetag{1.9cm}
\end{align}
where $x=p^2$, $y=q^2$ and $z=r^2$. The two-loop kernels $Q_3^0$ and $Q_4^0$ of Eqs.\ \eqref{3glooptrunc} and \eqref{4glooptrunc} are calculated in a straightforward way, but are more elaborate and therefore not explicitly given herein. 

The truncation is mainly motivated by its simplicity to achieve the two important properties that are multiplicative renormalizability and perturbative behavior. 
Nevertheless one would like to find a more theoretical justification for the truncation, which assumes that integrals with kernels containing specific combinations of full and bare vertices multiplied by various dressing functions can be well approximated by the same integrals over the tree-level kernels.

From the Slavnov-Taylor identities (STI) for the three-gluon, four-gluon and ghost-gluon vertices one observes that the vertices receive corrections which are related to the gluon and ghost dressing functions \cite{Ball:1980ax, Kim:1980ep, vonSmekal:1998is}. That this is not only true for the longitudinal parts of the vertices, constrained by these identities, but for the complete vertex can easily be seen from the principles of multiplicative renormalizability. The three-gluon vertex receives corrections proportional to $G/F$, the four-gluon vertex to $G^2/F$ and the ghost-gluon vertex to $G/G$, and the divergences of the vertices are given by the divergences of these ratios. 

All propagators and vertices in the Dyson-Schwinger equations \eqref{diagghost} and \eqref{diaggluon} are full, except for one vertex per diagram which is bare to avoid double counting in the integral equation. When briefly assuming factorizable \textit{Ans\"atze}\footnote{The factorizable \textit{Ansatz} replaces the full vertex by a product of bare vertex and dressing functions as suggested by the STIs, which correctly reproduces its divergence.} for the full vertices, a simple counting of dressing function corrections in each DSE diagram reveals that the MR truncation introduces an additional correction \textit{equivalent} to a dressing of the remaining bare vertex.  
Hence, the MR truncation could wrongly be interpreted as dressing all the vertices in the nonperturbative diagrams. That this is not true is easily understood from the following \textit{ab absurdo} argumentation. Placing full vertices on all interaction points would introduce double counting, and the perturbative results would not be reproduced. However, the MR truncation does reproduce the leading order perturbative result, hence refuting the allegation. As a matter of fact, the additional correction exactly ensures that the divergence of the integral matches that of the renormalization constant in the seed of the integral equations, thus yielding a finite renormalized dressing functions as required by the left hand sides of Eqs.\ \eqref{SD-Gh} and \eqref{SD-Gl}. 

It is known that the use of factorizable vertex Ans\"atze in the DSE does not reproduce the correct divergence and multiplicative renormalizability of the propagators. 
The exact dressed vertex has a structure that is much more complicated than just a multiplication of the bare vertex with ratios of dressing functions. Usually several tensor structures are involved, each having its own correction given by similar ratios of propagator dressings. 
This is well illustrated by the fermion-photon Curtis-Pennington (CP) vertex in quenched QED \cite{Curtis:1990zs}, constructed to satisfy the Ward-Takahashi identity and multiplicative renormalizability. 
The novel MR truncation assumes that the integrals satisfy specific cancellations {\it after} contractions and integrations. That this is not as far fetched as it at first seems can be seen from quenched QED. Substituting the quite elaborate CP-vertex in the quark DSE yields kernels of great simplicity \cite{Curtis:1991fb, Atkinson:1994mz} which are of the form suggested above. 
Furthermore, in \ref{App:QED} an analysis of the DSE for the massless quark propagator in quenched QED shows how the one-loop corrections to the various vertex structures contribute to the self-energy integral and combine to produce an MR dressing function, and how the same result can be obtained with a modified kernel which only uses the bare vertex structure.

Although the validity of the Ansatz could not yet be proven for QCD, the simplicity of the truncated equations, which satisfy multiplicative renormalizability, resummed leading order perturbation theory and take into account the two-loop nonperturbative diagrams, is a good enough reason to further study their properties. 

Now that the truncation scheme has been defined, the search for solutions of the coupled integral equations can be initiated. First, their infrared asymptotic behavior will be investigated analytically.

\section{Infrared Analysis}
\label{Sect:IR}

As shown in ref.\ \cite{Bloch:2001wz} a major feature of the MR truncation is that all diagrams in the gluon vacuum polarization in Eq.\ \eqref{diaggluon} contribute to the leading infrared behavior if the propagators obey infrared power laws. 
Indeed, in the infrared the propagator power laws combine to a fixed point for the running coupling, and as all diagrams in the MR truncation have kernels with similar dressing function structure up to an additional factor of $\alpha(q^2)$ they will all be infrared leading. 
This contrasts with the well-known fact that the sunset and squint diagrams are subleading for the ultraviolet logarithmic behavior. Also note that earlier truncations of the ghost-gluon system, with kernels violating multiplicative renormalizability, yield a very different infrared analysis where only the ghost loop contributes to the leading infrared behavior of the gluon vacuum polarization. 

The infrared analysis of the one-loop diagrams will be done in the same way as before (see ref.\ \cite{Bloch:2001wz} and references therein), and the method will be extended to compute the infrared contributions of the two-loop diagrams. 
As will be detailed further on, both gluon and ghost equations are individually satisfied by propagators obeying infrared power laws 
\be\label{powlaw0}
F_R(x) = A \, x^{\alpha} \,, \hspace{1cm} G_R(x) = B \, x^{\beta} \,,
\ee
where $x=p^2$.  When substituting these power laws in Eqs.\ \eqref{SD-Gh} and \eqref{SD-Gl}, the left and right hand sides of the integral equations yield power behaviors.
Requiring the leading infrared exponents on both sides to match, produces the constraint $\alpha = - 2\beta$, such that
\be\label{powlaw}
F_R(x) = A \, x^{2\kappa} \,, \hspace{1cm} G_R(x) = B \, x^{-\kappa} \,.
\ee
Because of Eq.~(\ref{rgi2}), these power laws generate an infrared fixed point for the running coupling
\be\label{nu}
\alpha_0 = \lim_{x \to 0} \alpha_\mu F_R(x) G_R^2(x) = \alpha_\mu A B^2 \to \text{constant} \,.
\ee

At this point it is useful to introduce the related function
\be\label{lambda}
\lambda(x) \equiv \frac{N_c \alpha(x)}{4\pi} =  \frac{N_c \alpha_\mu }{4\pi} F_R(x) G_R^2(x) \,.
\ee
From Eqs.\ \eqref{SD-Gh}, \eqref{SD-Gl}, \eqref{vacpoltot}, \eqref{SE-Ghtrunc}-\eqref{4glooptrunc} it is clear that the function $\lambda(x)$ only depends on its value $\lambda_\mu$ at the renormalization point, and is independent of the number of colors $N_c$ for the MR truncation presented herein\footnote{Note that this is also true up to third order in perturbation theory \cite{Tarasov:1980au}. At four-loop order $1/N_c^2$ corrections are generated \cite{vanRitbergen:1997va} due to non-vanishing contributions from non-planar diagrams.}. As will be shown in this section, the infrared value $\lambda_0$ is fixed by the coupled Dyson-Schwinger equations. Moreover, as $\lambda_0$ is independent of $N_c$, Eq.\ \eqref{lambda} shows that the infrared fixed point $\alpha_0$ of the running coupling is proportional to $1/N_c$.

Equating the coefficients of the leading power of $x$, for $x \to 0$, on both sides of each equation yields two relations between the infrared fixed point $\lambda_0$ and the leading infrared exponent $\kappa$: 
\be\label{gencondit}
f_G(\lambda_0, \kappa) = 0 \qquad , \qquad f_F(\lambda_0, \kappa) = 0 \,.
\ee
The existence of a consistent infrared power solution requires the gluon and ghost equation conditions (\ref{gencondit}) to be satisfied simultaneously. The solutions of Eq.\ \eqref{gencondit} will determine the values of the infrared fixed point $\lambda_0$ and exponent $\kappa$. The various loop integrals will now be computed and discussed.

\subsection{Ghost Equation}
\label{sect:ghost}

The leading power contributions to the ghost equation are calculated by substituting\footnote{\label{foot4}Even though the general solutions behave logarithmically at high momentum ($x \gg \LQCD^2$), power laws are used for all momenta in the infrared analysis. This procedure gives the correct leading power contribution because  pure power laws are themselves solutions of the coupled equations in the MR truncation, corresponding to $\LQCD \to \infty$. Ref.\ \cite{Atkinson:1998tu} showed that the leading infrared power is independent of $\LQCD$, which only enters the solution through the next term of its infrared asymptotic expansion.} 
the power laws (\ref{powlaw}) in Eq.\ (\ref{SD-Gh}) with self-energy (\ref{SE-Ghtrunc}):
\be\label{Ghpowlaw0}
\frac{x^\kappa}{B} = \tilde Z_3 - N_c g_\mu^2 A B \, \intq{q} \, T_0(x,y,z) \, y^{k_1} \, z^{k_2} \,,
\ee
where $x=p^2$, $y=q^2$, $z=(p-q)^2$, $\mathbf k = (k_1,k_2)$ is defined by $\mathbf k = \kappa(2 \mathbf b_\Sigma - \mathbf a_\Sigma)$, and $k_1+k_2=\kappa$ because $a_{\Sigma,1}+a_{\Sigma,2}=b_{\Sigma,1}+b_{\Sigma,2}=1$ (from Eqs.\ \eqref{Xm} and \eqref{SE-Ghtrunc}). Substituting the kernel $T_0$ given in \eqref{Txyz0} yields 
\be\label{Ghpowlaw}
x^\kappa = B \tilde Z_3 + N_c g_\mu^2 A B^2  \sum_{i,j,m}^{} c_{i j m} \, \intq{q} \, x^{i} \, y^{j+k_1} \, z^{m+k_2} \,,
\ee
where $i, j, m$ are integers satisfying $i+j+m=-2$, and the coefficients $c_{i j m}$ can be read off from Eq.\ (\ref{Txyz0}). The right hand side is a sum of integrals of the form
\be\label{integ-xyz}
\intq{q} \, x^\alpha \, y^\beta \, z^\gamma \,,
\ee
with $\alpha+\beta+\gamma = \kappa - 2$. In ref.\ \cite{Atkinson:1998zc} these integrals were solved using spherical coordinates, and the results were expressed in terms of generalized hypergeometric functions. A more concise, equivalent expression can be derived \cite{Lerche:2001} by noting that
integrals of type (\ref{integ-xyz}) are typical scalar integrals encountered when applying dimensional regularization in perturbative calculations and they are readily computed using Eq.\  (\ref{Iab}).

Equating the coefficients of the leading infrared power $x^\kappa$ on both sides of Eq.\ (\ref{Ghpowlaw}) gives
\be\label{chiGh}
\lambda_0 \; \chiG(k_1,k_2) = 1 \,,
\ee
with $\lambda_0=N_c \alpha_0/4\pi$ and $\alpha_0$ given by Eq.\ \eqref{nu}. The function $\chiG$ is given by the self-energy integral in Eq.\ \eqref{Ghpowlaw} which can be calculated analytically using Eq.\ (\ref{Iab}). Its general expression is
\be\label{chiGh-gen}
\chiG(k_1, k_2) = -\frac{3}{2}\frac{\Gamma(2+k_1)\Gamma(1+k_2)\Gamma(-k_1-k_2)}{\Gamma(1-k_1)\Gamma(2-k_2)\Gamma(3+k_1+k_2)} \,,
\ee
where the Gamma function recurrence relation was used repeatedly to bring the expression into its simplest form. The values of $k_1$ and $k_2$ depend on the details of the truncation through the sets $\mathbf a_\Sigma$ and $\mathbf b_\Sigma$ of \eqref{SE-Ghtrunc} which have not yet been fixed. 

For the ghost self-energy no symmetry between internal legs can be used to fix $\mathbf a_\Sigma$ and $\mathbf b_\Sigma$, and three different truncations are considered.
A natural choice corresponding to the bare ghost-gluon vertex approximation is obtained with $\mathbf a_\Sigma=(1,0)$ and $\mathbf b_\Sigma=(0,1)$, yielding a kernel proportional to $G_R(y) F_R(z) \stackrel{\text{IR}}{\sim} y^{-\kappa} z^{2\kappa}$. This choice gives $(k_1,k_2)=(-\kappa,2\kappa)$, and from Eq.\ \eqref{chiGh-gen}
\be\label{chi0}
\chiG^0(\kappa) \equiv \chiG(-\kappa, 2\kappa) = 
\frac{6 \,\Gamma^2(2-\kappa)\Gamma(2\kappa)}{\Gamma(3-2\kappa)\Gamma(3+\kappa)\Gamma(1+\kappa)} \,.
\ee
An expression equivalent to Eq.\ (\ref{chi0}) was first given in ref.\ \cite{Atkinson:1998zc}, where it was written in terms of generalized hypergeometric functions. 

The $G/G$ nature of the ghost-gluon vertex corrections makes it possible for the bare vertex to satisfy MR and reproduce PT. 
Nevertheless, it is clear that the exact dressed ghost-gluon vertex is not identical to the bare one, as can for example be seen from its Slavnov-Taylor identity \cite{vonSmekal:1998is}. That identity suggests vertex corrections like $G(r^2)/G(q^2)$ where $q$ is the ghost momentum in the loop, and $r$ the gluon momentum. 
Approximating the full ghost-gluon vertex by a bare vertex multiplied by such a correction leads to the truncation $\mathbf a_\Sigma=(0,1)$ and $\mathbf b_\Sigma=(0,1)$, and the kernel is proportional to $G_R(z) F_R(z)$ with infrared limit  $z^{\kappa}$, i.e.  $k_1=0$, $k_2=\kappa$. In this case
\be\label{chi-1}
\chiG^V(\kappa) \equiv \chiG(0,\kappa) = \frac{3}{2\kappa(1+\kappa)(2+\kappa)(1-\kappa)} \,.
\ee

However, as explained in the previous section, in general the MR truncation requires an additional vertex type correction to satisfy MR and reproduce the correct perturbative behavior of the integrals. This is not obvious for the ghost self-energy because of the $G/G$ nature of the vertex corrections, but is clear for diagrams containing three-gluon, four-gluon and quark-gluon vertices. Introducing such an additional correction leads to another variant, which is very similar to the multiplicatively renormalizable truncation of the quark equation \cite{Bloch:2002eq}. This corresponds to $\mathbf a_\Sigma=(-1,2)$ and $\mathbf b_\Sigma=(0,1)$, yielding a kernel proportional to $\alpha(z)/G_R(y)$ (using Eq.\ (\ref{rgi2})) with infrared limit  $y^{\kappa}$, i.e.  $k_1=\kappa$, $k_2=0$, and
\be\label{chi-a}
\chiG^\alpha(\kappa) \equiv \chiG(\kappa, 0) = \frac{3}{2\kappa(2+\kappa)} \,.
\ee

\begin{figure}[ht]
\begin{center}
\includegraphics[width=11.0cm]{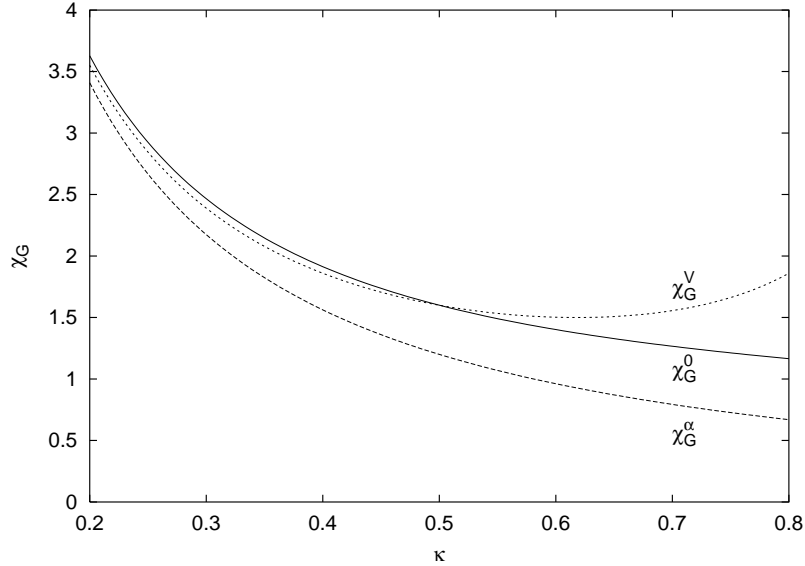}
\end{center}
\vspace{-15pt}
\caption{\label{a0ghost}Variation of the infrared ghost self-energy integral $\chiG$ as function of the infrared exponent $\kappa$ for three specific choices of the kernel in Eq.\ (\ref{SE-Ghtrunc}): $\chiG^0$ (\ref{chi0}) using a bare ghost-gluon vertex, $\chiG^V$ \eqref{chi-1} with a factorizable vertex Ansatz, and $\chiG^\alpha$ (\ref{chi-a}) for a $\alpha(z)/G_R(y)$ kernel.}
\end{figure}

The three truncations \eqref{chi0}, \eqref{chi-1} and \eqref{chi-a} will respectively be referred to as bare truncation, V-truncation and $\alpha$-truncation.
The variation of $\chiG$ as function of $\kappa$ for these three cases is shown in Fig. \ref{a0ghost}. The plotted quantity is inversely proportional to the infrared fixed point of the running coupling (see Eq.\ \eqref{chiGh}), and from the figure one can get information about the dependence of the fixed point on $\kappa$ using different kernel corrections. For instance, a value of $\kappa=0.5$ would correspond to a fixed point value $\lambda_0=5/8$ for the bare truncation and V-truncation, and $\lambda_0=5/6$ for the $\alpha$-truncation. Global infrared and ultraviolet convergence criteria of the integrals require $0 < \kappa < 2$. 

\subsection{Gluon Equation}\label{Sec:GlEq}

Next, the various contributions to the gluon vacuum polarization are investigated. It is important to note that, in contrast to the studies \cite{vonSmekal:1998is, Atkinson:1998tu, Atkinson:1998zc}, all the vacuum polarization diagrams (\ref{pightrunc}) to (\ref{4glooptrunc}) of the MR truncation will be shown to contribute to the leading order infrared power of the gluon equation (\ref{SD-Gl}). The infrared analysis yields the following equation after substituting the power laws \eqref{powlaw} in Eq.\ (\ref{SD-Gl})
\be
\frac{x^{-2\kappa}}{A} = Z_3 - \lim_{x\to0} [ \Pi^{2c}(x) + \Pi^{2g}(x) + \Pi^{3g}(x) + \Pi^{4g}(x) ] \,,
\ee
or for the coefficients of the leading power in $x$:
\be\label{IR-Gl}
1 = - \lim_{x\to 0} A x^{2\kappa} \left[ \Pi^{2c}(x) + \Pi^{2g}(x) + \Pi^{3g}(x)+\Pi^{4g}(x) \right] \,.
\ee

Each contribution is now investigated in turn. The infrared ghost loop contribution (\ref{pightrunc}) to Eq.\ \eqref{IR-Gl} is given by
\be\label{Glghpowlaw}
\lambda_0 \, \chiF^{2c}(k_1,k_2) \equiv - \lim_{x\to 0} A x^{2\kappa} \Pi^{2c}(x) = N_c \, g_\mu^2 A B^2  \sum_{i,j,m} c_{i j m} \intq{q} \, x^{i+2\kappa} y^{j+k_1} z^{m+k_2} 
\ee
with $x=p^2$, $y=q^2$, $z=(p-q)^2$. The integers $i, j, m$ satisfy $i+j+m=-2$, and the coefficients $c_{i j m}$ can be read off from Eq.\ (\ref{Mxyz0}). Furthermore, $\mathbf k = (k_1,k_2)$ is defined by $\mathbf k = \kappa(2 \mathbf b_{2c} - \mathbf a_{2c})$, and $k_1+k_2=-2\kappa$ because $a_{2c,1}+a_{2c,2}=2$, $b_{2c,1}+b_{2c,2}=0$ (from Eq.\ \eqref{pightrunc}). 

Although one can easily calculate this contribution for arbitrary $\mathbf a_{2c}$ and $\mathbf b_{2c}$,  their values are naturally fixed by symmetry considerations between the two internal ghost legs: $\mathbf a_{2c}=(1,1)$, $\mathbf b_{2c}=(0,0)$. The kernel is then proportional to $G_R(y) G_R(z)$, and its infrared contribution, for which $\mathbf k=(-\kappa,-\kappa)$, is
\be\label{chi-cc}
\chiF^{2c}(\kappa) = \frac{\Gamma^2(2-\kappa)\Gamma(2\kappa)}{\Gamma(4-2\kappa)\Gamma^2(1+\kappa)} \,.
\ee
This choice corresponds to an \textit{effective} bare ghost-gluon vertex approximation. Although one could argue that different ghost-gluon vertex approximations are used in the ghost (Sect. \ref{sect:ghost}) and gluon equations, this is not an inconsistency as the truncation considers effective kernel corrections rather than Ans\"atze for the full dressed vertices. Because of the different tensor structures of the ghost self-energy and gluon vacuum polarization diagrams, kernel corrections can be of different natures for both equations.
Furthermore, for the ghost loop contribution to the gluon vacuum polarization, corrections analogous to those of Sect. \ref{sect:ghost} would depend on the external gluon momentum, and are not allowed in the effective kernel truncations developed in this paper as they would destroy the perturbative properties of the truncation. 

\begin{figure}[ht]
\includegraphics[width=7.8cm]{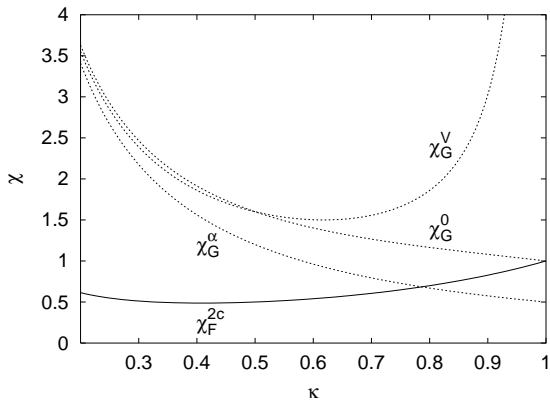}
\parbox[b]{5.5cm}{\caption{\label{chi1}Variation of the infrared ghost loop integral $\chiF^{2c}$ (\ref{chi-cc}) in the gluon vacuum polarization, together with the ghost self-energy integrals $\chiG^0$ (\ref{chi0}), $\chiG^V$ (\ref{chi-1}) and $\chiG^\alpha$ (\ref{chi-a}) from Fig. \ref{a0ghost}, as function of the infrared exponent $\kappa$. The crossing of $\chiF^{2c}$ and $\chi^0_G$ at $\kappa=1$ was suggested as a possible solution for the equations in the leading ghost-loop truncation of ref.\ \cite{Atkinson:1998zc}.}}
\end{figure}

The evolution of $\chiF^{2c}$ is shown in Fig.\ \ref{chi1} together with the ghost curves for $\chiG^0$, $\chiG^V$ and $\chiG^\alpha$. In the leading ghost-loop truncation of ref.\ \cite{Atkinson:1998zc} the consistency condition (\ref{gencondit}) between ghost and gluon equations is $1/\lambda_0 = \chiG(\kappa) = \chiF^{2c}(\kappa)$, and the crossing of $\chiF^{2c}$ and $\chi^0_G$ at $\kappa=1$, with $\lambda_0=1$, was suggested as a possible solution of the coupled equations. Ref.\ \cite{Atkinson:1998zc} only described the infrared analysis, and the numerical study of the equations did not reveal solutions satisfying the equations over the whole momentum range. 
Using the $\alpha$-truncation (\ref{chi-a}) of the ghost equation would yield a slightly different infrared solution in the leading ghost-loop approximation, with $\kappa=0.79$ and a relatively large coupling $\lambda_0=1.46$. Note that the results can be sensitive to the truncation as can be seen from the V-truncation for which no solution for $\kappa$ exists, as no intersection between this ghost curve and the gluon curve occurs in Fig. \ref{chi1}. 

The full numerical investigation of the leading-ghost loop truncation was repeated in the current study and it confirms that no solution can be found connecting the infrared to the ultraviolet asymptotic behavior. The failure to find numerical solutions does not prove that such do not exist, as the {\it numerical} solution of the equations could be sensitive to instabilities. However, as will be shown later on, the numerical method applied to solve the equations is quite powerful, and it is therefore very 
likely that this truncation has no solution at all, and that the infrared solutions are spurious solutions of the infrared analysis.
This problem will be resolved in an elegant manner by the MR truncation.

In a further attempt to find numerical solutions of the coupled equations in the leading ghost-loop approximation with exact angular integrals, refs.\ \cite{Lerche:2001, Lerche:2002ep, Fischer:2002hn} extracted the gluon dressing function from the tensor equation \eqref{SD-Gl-full} using ${\cal P}_{\mu\nu}$ of Eq.\ \eqref{project} with \textit{arbitrary} $n$. They reported new infrared solutions with $\kappa \gtrapprox 0.5$ for $n \neq 4$. These solutions are spurious as is explained below.

\begin{figure}[ht]
\begin{center}\includegraphics[width=11.0cm]{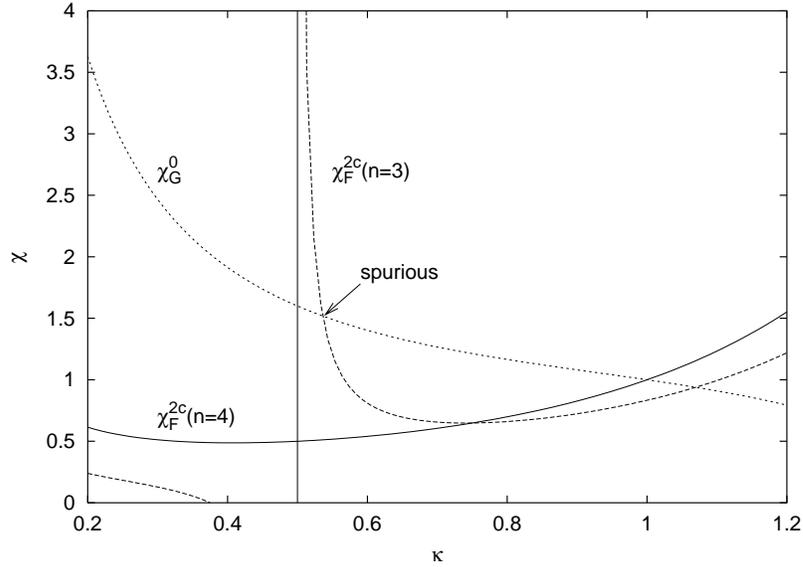}\end{center}
\vspace{-15pt}
\caption{\label{chi1b}Variation of the infrared ghost loop integral $\chiF^{2c}$ (\ref{chi-ccn}) in the gluon vacuum polarization for $n=4$ and $n=3$, together with the ghost self-energy integral $\chiG^0$ (\ref{chi0}), as function of the infrared exponent $\kappa$. The additional crossing for the $n=3$ curve is spurious as it is due to the divergence of $\chiF^{2c}$ at $\kappa=0.5$, which is related to the presence of spurious ultraviolet quadratic divergences when $n \neq 4$ in Eq.\ (\ref{project}).}
\end{figure}

Although the solution can depend on $n$ if the DSE truncation violates the transversallity of the gluon vacuum polarization, one hopes that its sensitivity is minimal. 
However, an additional problem occurs when using a non-BP tensor ($n \neq 4$) as spurious quadratic divergences are generated in the gluon vacuum polarization \cite{Brown:1989bn}. These divergences have to be removed properly in order to get renormalizable solutions. Performing the infrared analysis of the ghost loop in the gluon vacuum polarization for arbitrary $n$ leads to an extension of Eq.\ \eqref{chi-cc}
\begin{align}\label{chi-ccn}
\chiF^{2c}(n,\kappa) &= \left[ n\,\left( -3 + 4\,\kappa  \right) + 6 - 4\,\kappa  \right] \frac{ {\Gamma^2(2 - \kappa )}\,\Gamma(-1 + 2\,\kappa )}{6\,\Gamma(4 - 2\,\kappa )\,{\Gamma^2(1 + \kappa )}} \notag\\
    &= \chiF^{2c}(4,\kappa) +
    \underline{\left( n-4 \right)} \,\left( -3 + 4\,\kappa  \right) \frac{{\Gamma^2(2 - \kappa )}\,\Gamma(-1 + 2\,\kappa )}{6\,\Gamma(4 - 2\,\kappa )\,{\Gamma^2(1 + \kappa )}} \,,
\end{align}
where $\chiF^{2c}(4,\kappa)$ is given by Eq.\ (\ref{chi-cc}). Equation \eqref{chi-ccn} shows how $\chiF^{2c}(n,\kappa)$ can be written as a sum of two terms: one regular and the other singular at $\kappa=0.5$. It is easy to check that the singular factor $\Gamma(-1+2\kappa)$ is exactly generated by the terms of $M_0$ (extension of Eq.\ \eqref{Mxyz0} for arbitrary $n$) which are quadratically divergent in the ultraviolet limit. The singularity only occurs for $n \neq 4$ and is spurious.  
Figure \ref{chi1b} shows $\chiF^{2c}$ for $n=4$ and $n=3$, together with the ghost curve $\chi_G^0$. The singularity at $\kappa=0.5$ for $n \neq 4$ inevitably introduces an additional, {\it spurious} crossing of the ghost and gluon curves just above $\kappa=0.5$. 

When removing the kernel terms responsible for the quadratic divergence in a natural way, the singular Gamma function in Eq.\ \eqref{chi-ccn} disappears and no spurious solution for $\kappa$ is found. The only solution left for $n \neq 4$ is the one that continuously connects to the $\kappa=1$ solution for $n=4$ \cite{Atkinson:1998zc}. Unfortunately, after removal of the quadratically divergent terms the integral equations do not have overall solutions, as was the case for $n=4$ in ref.\ \cite{Atkinson:1998zc}, and as could be expected the absence of solutions is independent of $n$.

The absence of numerical solutions is remedied in ref.\ \cite{Fischer:2002hn} by introducing a specific procedure to remove the quadratic divergences for $n \neq 4$.
The terms responsible for the quadratic divergences are not simply removed from the respective ghost and gluon loop kernels where they originally occur. Rather, the subtraction in the gluon loop is adapted so that it cancels the ultraviolet quadratic divergences coming from both gluon and ghost loops, while the ghost loop is left unsubtracted. As the gluon loop is infrared \textit{subleading} in that truncation, the subtraction scheme leaves the leading infrared analysis of Eq.\ \eqref{chi-ccn} unchanged. The truncation yields infrared solutions with $\kappa \gtrapprox 0.5$ for $n \neq 4$, as indicated in Fig.\ \ref{chi1b}, and is free of ultraviolet quadratic divergences at the same time. Furthermore, the numerical treatment found overall solutions for the integral equations, connecting the asymptotic behaviors at low and high momenta. 

This truncation scheme was the first one to yield {\it full} numerical solutions in a treatment with exact angular integrals (for any $n$, except $n=4$). Nevertheless, it is difficult to evaluate the usefulness of the truncation because of the peculiar scheme used to subtract the quadratic divergences. Clearly, this subtraction scheme can only be used as long as the gluon loop does not contribute to leading infrared order, and this is in contradiction with the MR truncation of Sect.\ \ref{Sect:Truncation}. The goal herein is to go beyond that truncation, avoiding quadratic divergences from the very beginning by using $n=4$, and to develop a truncation respecting multiplicatively renormalizability, for which all diagrams contribute to leading infrared order.

As was shown in ref. \cite{Bloch:2001wz} the gluon loop also contributes to the leading infrared power behavior of the gluon vacuum polarization in the MR truncation. The treatment is analogous to that of the ghost loop, and its contribution is given by
\be\label{Glglpowlaw}
\lambda_0 \, \chiF^{2g}(k_1,k_2) \equiv - \lim_{x\to 0} A x^{2\kappa} \Pi^{2g}(x) \,,
\ee
with $\mathbf k = (k_1,k_2)$ defined by $\mathbf k = \kappa(2 \mathbf b_{2g} - \mathbf a_{2g})$, and $k_1+k_2=-2\kappa$ because $a_{2g,1}+a_{2g,2}=2$, $b_{2g,1}+b_{2g,2}=0$. This leads to a sum of integrals similar to that of Eq.\ (\ref{Glghpowlaw}), where the coefficients $c_{ijm}$ are now determined by the kernel (\ref{Qxyz0}).

The values $\mathbf a_{2g}$ and $\mathbf b_{2g}$ are again naturally fixed based on symmetry considerations between the two internal gluon legs: $\mathbf a_{2g}=(1,1)$, $\mathbf b_{2g}=(0,0)$. The kernel is proportional to $G_R(y) G_R(z)$, and its infrared contribution is\footnote{The expressions $\chiF^{2c}$ \eqref{chi-cc} and $\chiF^{2g}$ \eqref{chi-2g} given here in terms of Gamma functions are identical to those expressed in terms of generalized hypergeometric functions in ref.\ \cite{Bloch:2001wz}, as was checked numerically.}
\be\label{chi-2g}
\chiF^{2g}(\kappa) = \left( 15\,k^2 - 4\,k - 14 \right) \frac{\Gamma(1 - k)\,
    \Gamma(2 - k)\,\Gamma(2\,k)}{\Gamma(4 - 2\,k)\,{\Gamma^2(2 + k)}} \,.
\ee
Note that in this case the effective MR kernel automatically corresponds to some implicitly defined dressed vertex. Using a bare triple gluon vertex in $Q$ of Eq.\ \eqref{pigl} would yield equations that do not satisfy Eq.\ \eqref{pigltrunc}, would violate multiplicative renormalizability and would furthermore not contribute to the leading infrared power. 

\begin{figure}[ht]
\includegraphics[width=8.0cm]{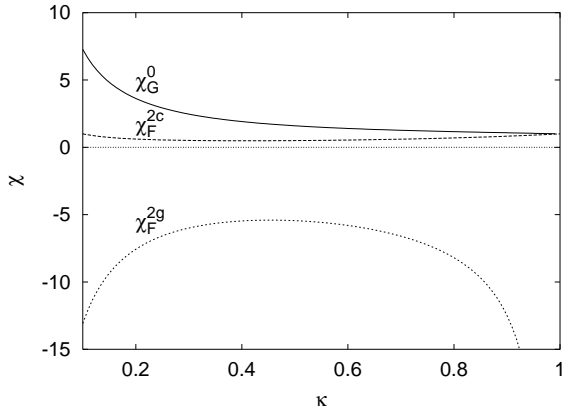}
\parbox[b]{5.3cm}{\caption{\label{chi2}Contributions $\chiG^0$, $\chiF^{2c}$ and  $\chiF^{2g}$ as function of the infrared exponent $\kappa$. The ghost curve $\chiG^0$ and the gluon curve $\chiF^{2c}+\chiF^{2g}$ do not intersect, and power laws are not solutions in the \textit{one-loop-only} MR truncation.}}
\end{figure}

When neglecting the two-loop diagrams and only taking into account the one-loop contributions in the gluon vacuum polarization, the infrared constraint \eqref{IR-Gl} from the gluon equation is
\be\label{chiGl}
\lambda_0 \; [\chiF^{2c}(\kappa) + \chiF^{2g}(\kappa)] = 1 \,.
\ee
The existence of consistent infrared power solutions requires the ghost and gluon relations (\ref{chiGh}) and (\ref{chiGl}) to be satisfied simultaneously:
\be\label{consist}
\chiG(\kappa) = \chiF^{2c}(\kappa) + \chiF^{2g}(\kappa) \,.
\ee
Solutions of this equation would yield possible values for the leading infrared exponent $\kappa$. The corresponding $\alpha_0$ would then be computed from Eq.~(\ref{chiGh}) or (\ref{chiGl}).
In Fig.\ \ref{chi2} the gluon loop contribution $\chiF^{2g}$ is drawn together with $\chiF^{2c}$ and $\chiG$, and it is clear that the condition (\ref{consist}) cannot be satisfied for any $\kappa$ between 0 and 1 because of the large negative contribution of the gluon loop. This lead to the conclusion in ref.\ \cite{Bloch:2001wz} that the MR \textit{one-loop-only} truncation, where only the ghost and gluon loops are retained in the gluon vacuum polarization, does not support infrared power laws for the propagators. 

The absence of power law solutions in that truncation could be interpreted as a setback, but it was suggested in ref.\ \cite{Bloch:2001wz} that taking into account the two-loop diagrams in the DSEs could substantially change the infrared behavior of the equations. Herein it will be shown that the absence of consistent infrared power law solutions is an artifact of an incomplete truncation due to the omission of the two-loop diagrams, and that taking these into account, albeit in a simple realization, will reinstate power laws in a very natural way. 

The following section will illustrate how the infrared analysis of the two-loop gluon vacuum polarization diagrams can be performed quite elegantly.

\subsection{Two-loop Diagrams}
\label{Sect:Two-loop}

The sunset contribution (\ref{3glooptrunc}) to the gluon vacuum polarization in the MR truncation is given by
\be\label{3gloopc}
\Pi^{3g}(x) = N_c^2 g_\mu^4 \intq{q_1} \intq{q_2} \, Q_3^0(p,q_1,q_2,q_3) \, G_R^4[\mathbf a_{3g}] \, F_R^1[\mathbf b_{3g}] \,.
\ee

Taking the infrared limit yields the contribution of the sunset diagram to the coefficient of the leading infrared power $x^{-2\kappa}$ in the gluon equation \eqref{IR-Gl}:
\be\label{3gloop-ir}
\lambda_0^2 \, \chiF^{3g}(k_1,k_2,k_3) \equiv  - \lim_{x\to 0} A x^{2\kappa} \Pi^{3g}(x) \,.
\ee
Substituting the power laws \eqref{powlaw} yields\footnote{Power laws are assumed for all momenta as explained in footnote \ref{foot4} on page \pageref{foot4}.}
\be\label{3gloop-ir2}
\chiF^{3g}(k_1,k_2,k_3) = - (16\pi^2)^2  \intq{q_1} \intq{q_2} \, Q_3^0(p,q_1,q_2,q_3) \, x^{2\kappa} z_1^{k_1} z_2^{k_2} z_3^{k_3}\,,
\ee
with $x=p^2$, $z_i=q_i^2$. Furthermore, $\mathbf k = (k_1,k_2,k_3)$ is defined by $\mathbf k = \kappa(2 \mathbf b_{3g} - \mathbf a_{3g})$, and $k_1+k_2+k_3=-2\kappa$ because $a_{3g,1}+a_{3g,2}+a_{3g,3}=4$, $b_{3g,1}+b_{3g,2}+b_{3g,3}=1$.

To compute the sunset diagram it is crucial to note that it can be considered as a sequence of two single loop integrations. First, the $(q_2,q_3)$-loop, enclosed in the box in the diagram shown below, is integrated out, and then the remaining $(q_1, q_s)$-loop, with $q_s=q_2+q_3=p-q_1$, is computed.
\begin{align}\begin{split}
\includegraphics[width=2.5cm]{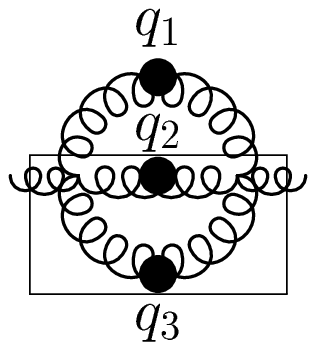}\label{3gluoninteg}
\end{split}\end{align}
The integration rule (\ref{Iab}) cannot be applied as such to compute the inner loop, because the kernel $Q_3$ cannot be written as a function of squared momenta only, as could be done for the one-loop diagrams. There is always one angle that cannot be eliminated. Fortunately, the integral can still be solved in a fairly easy way by treating it as a sum of tensor integrals of rank 4 or lower. 

The various tensor terms in the kernel of Eq.\ (\ref{3gloop-ir2}) have denominators proportional to $q_1^{2 a_1} q_2^{2 a_2} q_3^{2 a_3}$, with non-integer powers $a_1, a_2, a_3$. To transform this into known integrals, first the Feynman trick (\ref{FT1}) is applied to rewrite the denominators of the inner loop integral as
\be
\frac{1}{q_2^{2 a_2} q_3^{2 a_3}} =  \frac{\Gamma(a_2+a_3)}{\Gamma(a_2)\Gamma(a_3)} \int_0^1 dt \;  \frac{t^{a_2-1} (1-t)^{a_3-1} }{\left[ t q_2^2 + (1-t) q_3^2 \right]^{a_2+a_3}}\,.
\ee
The momentum $q_3$ is eliminated by introducing the outer loop momentum $q_s = q_2+q_3$
\be
\frac{1}{q_2^{2 a_2} q_3^{2 a_3}} = \frac{\Gamma(a_2+a_3)}{\Gamma(a_2)\Gamma(a_3)} \int_0^1 dt \; 
\frac{t^{a_2-1} (1-t)^{a_3-1}}{\left[ q_2^2-2(1-t) q_2 \cdot q_s  + (1-t) q_s^2 \right]^{a_2+a_3}} \,.
\ee
To eliminate the dot product in the denominator, so that it is written in terms of squared momenta only, the variable transformation $\ell = q_2 - (1-t)q_s$ is introduced:
\be
\frac{1}{q_2^{2 a_2} q_3^{2 a_3}} =   \frac{\Gamma(a_2+a_3)}{\Gamma(a_2)\Gamma(a_3)} \int_0^1 dt \; \frac{t^{a_2-1} (1-t)^{a_3-1}}{\left[ \ell^2+t(1-t)q_s^2 \right]^{a_2+a_3}} \,.
\ee
The inverse variable transformations
\be
q_2 = \ell + (1-t) q_s \quad , \qquad q_3 = t q_s - \ell \,,
\ee
are then used to replace the four-momenta components of $q_2$ and $q_3$ by $\ell$ and $q_s$ in the numerators of the tensor integrals of Eq.\ (\ref{3gloop-ir2}). The ensuing inner loop integrals can be calculated using the tensor integral formulae Eqs.\ (\ref{J0}) to (\ref{J6}). Consequently, the remaining integration over the Feynman parameter $t$ is easily performed using Eq.\ (\ref{integt}). After integration of the inner loop the kernel of the external $(q_1,q_s)$ loop integral is a function of $p$, $q_1$ and $q_s$ which can be written as a function of squared momenta only, and this integral can be computed using Eq.\ (\ref{Iab}). This complete procedure was programmed using the symbolic algebra program Form \cite{Vermaseren:2000nd} and yields the following result for the integral (\ref{3gloop-ir2}) with $Q_3^0$ defined as the tree-level version of kernel (\ref{kernelQ3}):
\be\label{chi3g}
\begin{split}
\chiF^{3g}(k_1,k_2,k_3) &= \frac{3}{8}\left( 1 + {k_1} + {k_2}  + {k_3} - 2\,{k_1}\,{k_2} - 2\,{k_1}\,{k_3} - 2\,{k_2}\,{k_3} - 8\,{k_1}\,{k_2}\,{k_3} \right) \\
& \times \frac{\Gamma(1 + {k_1})\,\Gamma(1 + {k_2})\,\Gamma(1 + {k_3})\,\Gamma(-{k_1} - {k_2} - {k_3}) }
{\Gamma(2 - {k_1})\,\Gamma(2 - {k_2})\,\Gamma(2 - {k_3})\,\Gamma(4 + {k_1} + {k_2} + {k_3})} \,.
\end{split}\raisetag{20pt}
\ee
As expected from diagram \eqref{3gluoninteg}, this result is invariant under $k_1 \leftrightarrow k_2 \leftrightarrow k_3$. The parameters in the MR truncation of the sunset diagram (\ref{3gloopc}) are chosen to respect the symmetry between the three internal gluon legs, i.e. $\mathbf a_{3g}=(4/3,4/3,4/3)$ and $\mathbf b_{3g}=(1/3,1/3,1/3)$ corresponding to a kernel 
\be
Q_3^0(p,q_1,q_2,q_3) \; F_R^{1/3}(z_1) G_R^{4/3}(z_1) F_R^{1/3}(z_2) G_R^{4/3}(z_2) F_R^{1/3}(z_3) G_R^{4/3}(z_3) \,.
\ee
Hence, the set $\mathbf k$ in the infrared equation (\ref{3gloop-ir2}) respects this symmetry too, and $k_1=k_2=k_3=-2\kappa/3$. For such a choice, Eq.\ \eqref{chi3g} becomes 
\begin{align}
\chiF^{3g}(\kappa) &\equiv 
\chiF^{3g}(-2\kappa/3,-2\kappa/3,-2\kappa/3)
\notag\\
\label{chi3gb}
&= \frac{3}{8}\left(1-2\kappa-\frac{8}{3}\kappa^2+\frac{64}{27}\kappa^3\right) \frac{\Gamma^3(1-\frac{2\kappa}{3})\Gamma(2\kappa)}{\Gamma^3(2+\frac{2\kappa}{3})\Gamma(4-2\kappa)} \,.
\end{align}

The treatment of the squint diagram can be performed in a way analogous to that of the sunset diagram. The squint contribution (\ref{4glooptrunc}) to the gluon vacuum polarization in the MR truncation is
\be\label{4gloopc}
\Pi^{4g}(x) = N_c^2 g_\mu^4  \intq{q_1} \intq{q_2} \, Q_4^0(p,q_1,q_2,q_3,q_4) \, G_R^4[\mathbf a_{4g}] \,  F_R^1[\mathbf b_{4g}] \,.
\ee

Again, the contribution to the coefficient of the infrared leading power is given by taking the infrared limit,
\be\label{4gloop-ir}
\lambda_0^2 \, \chiF^{4g}(k_1,k_2,k_3,k_4) \equiv - \lim_{x \to 0} A x^{2\kappa} \Pi^{4g}(x) \,,
\ee
and substituting the infrared power laws:
\be
\chiF^{4g}(k_1,k_2,k_3,k_4) = -(16\pi^2)^2  \intq{q_1} \intq{q_2} \, Q_4^0(p,q_1,q_2,q_3,q_4) \, x^{2\kappa} z_1^{k_1} z_2^{k_2} z_3^{k_3} z_4^{k_4} \,,
\ee
with $x=p^2$, $z_i=q_i^2$. Furthermore, $\mathbf k = (k_1,k_2,k_3,k_4)$ is defined by $\mathbf k = \kappa(2 \mathbf b_{4g} - \mathbf a_{4g})$, and $k_1+k_2+k_3+k_4=-2\kappa$ because $a_{4g,1}+a_{4g,2}+a_{4g,3}+a_{4g,4}=4$, $b_{4g,1}+b_{4g,2}+b_{4g,3}+b_{4g,4}=1$.
 
As before, the two-loop integral can be computed by evaluating two consecutive one-loop integrals: first the inner-loop integral over $(q_2,q_3)$ and then the remaining outer-loop integral over $(q_1,q_4)$ as shown in the diagram below.
\begin{align}
\begin{split}
\includegraphics[width=2.5cm]{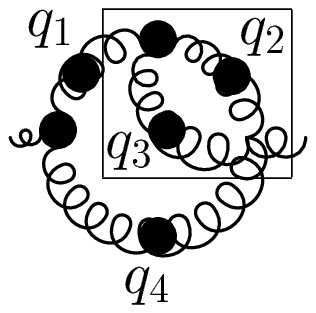}\label{4gluoninteg}
\end{split}
\end{align}
The steps to be followed are exactly the same as those described above for the sunset diagram. The inner loop integral over $(q_2,q_3)$ can be transformed into known tensor integrals of rank 5 or lower. Once these have been integrated out, the remaining one-loop integral has a kernel which is a function of $p, q_1, q_4$ which can easily be solved using Eq.\ (\ref{Iab}). The final result is
\be\label{chi4g}
\begin{split}
\chiF^{4g}(k_1,k_2,k_3,&k_4) = -\frac{3}{8}\left(  3\,{{k_2}}^2+ {k_2}\,{k_3} + 3\,{{k_3}}^2 + {k_2} + {k_3} -5  \right) \\
& \times \left( 13 + 3 \, \left({k_1} + {k_2} + {k_3}\right) + 13\,{k_4} + 6\,\left( {k_1} + {k_2} + {k_3} \right) \,{k_4} - 6\,{{k_4}}^2 \right) \\
& \times \frac{\Gamma(1 + {k_2})\,\Gamma(1 + {k_3})\,\Gamma(-{k_2} - {k_3})}
{\Gamma(2 - {k_2})\,\Gamma(2 - {k_3})\,\Gamma(3 + {k_2} + {k_3})\,} \\ 
& \times \frac{\Gamma(1 + {k_4})\,\,\Gamma(2 + {k_1} + {k_2} + {k_3})\,\Gamma(-{k_1} - {k_2} - {k_3} - {k_4})}{\Gamma(2 - {k_4})\,\Gamma(1 - {k_1} - {k_2} - {k_3})\,\Gamma(4 + {k_1} + {k_2} + {k_3} + {k_4})}\,.
\end{split}\raisetag{40pt}
\ee
Note that $\chiF^{4g}$ is symmetric under interchange of $q_2$ and $q_3$ as is expected from diagram \eqref{4gluoninteg}. The value of $\chiF^{4g}$ depends on the choice of $\mathbf a_{4g}$ and $\mathbf b_{4g}$ in Eq.\ \eqref{4glooptrunc}. 

In the next section, the existence of infrared power solutions will be investigated taking into account all the contributions calculated above.

\subsection{Consistency Condition}
\label{Sect:consist}

The infrared consistency condition (\ref{consist}) has to be modified when the two-loop diagrams are taken into account in the gluon vacuum polarization. For the gluon equation the relation between the infrared fixed point of the coupling and the infrared exponent of the propagator power laws can be derived from Eq.\ (\ref{IR-Gl}) using Eqs.\ \eqref{Glghpowlaw}, \eqref{Glglpowlaw}, \eqref{3gloop-ir} and \eqref{4gloop-ir}:
\be\label{eql0}
1 = \lambda_0 \, \chiF^{1\ell}(\kappa) + \lambda_0^2 \, \chiF^{2\ell}(\kappa) \,,
\ee
where the one-loop and two-loop contributions are defined by\footnote{The $\chi$'s depend on $\mathbf k=\kappa(2\mathbf b - \mathbf a)$, and can be written as functions of $\kappa$ for fixed $\mathbf a$'s and $\mathbf b$'s.}
\be
\chiF^{1\ell}(\kappa) = \chiF^{2c}(\kappa) + \chiF^{2g}(\kappa) \quad , \quad
\chiF^{2\ell}(\kappa) = \chiF^{3g}(\kappa)+\chiF^{4g}(\kappa) \,.
\ee
Equation \eqref{eql0} is solved by
\be\label{chiGl2}
\frac{1}{\lambda_0} = \frac{1}{2}\left( \chiF^{1\ell}(\kappa) \pm \sqrt{\left[\chiF^{1\ell}(\kappa)\right]^2 + 4 \,\chiF^{2\ell}(\kappa)} \right) \,.
\ee
As discussed in Sect. \ref{Sec:GlEq}, the one-loop contribution is negative, and therefore a positive solution for $\lambda_0$ can only occur for the "+"-solution in Eq.\ \eqref{chiGl2}.

The infrared consistency condition between ghost and gluon equations simply follows by requiring Eqs.\ (\ref{chiGh}) and (\ref{chiGl2}) to be satisfied simultaneously, and the infrared exponent $\kappa$ is determined from
\be\label{consist2}
\chiG(\kappa) = \frac{1}{2}\left( \chiF^{1\ell}(\kappa) + \sqrt{\left[\chiF^{1\ell}(\kappa)\right]^2 + 4 \,\chiF^{2\ell}(\kappa)} \right)  \,.
\ee

For the simple MR kernel truncations developed in this paper the parametrization of the ghost loop \eqref{pightrunc}, gluon loop \eqref{pigltrunc} and sunset diagram \eqref{3glooptrunc} were fixed by symmetry considerations between the internal momenta. However, the parametrization of the squint diagram \eqref{4glooptrunc} has not yet been fixed as $\mathbf a_{4g}, \mathbf b_{4g}$ still have to be assigned specific values. A major concern of this study is to investigate if sets can be found for which the infrared consistency condition (\ref{consist2}) is satisfied, and if so, if numerical solutions for the propagators exist having this asymptotic infrared behavior. 

\begin{figure}[b]
\includegraphics[width=6.5cm]{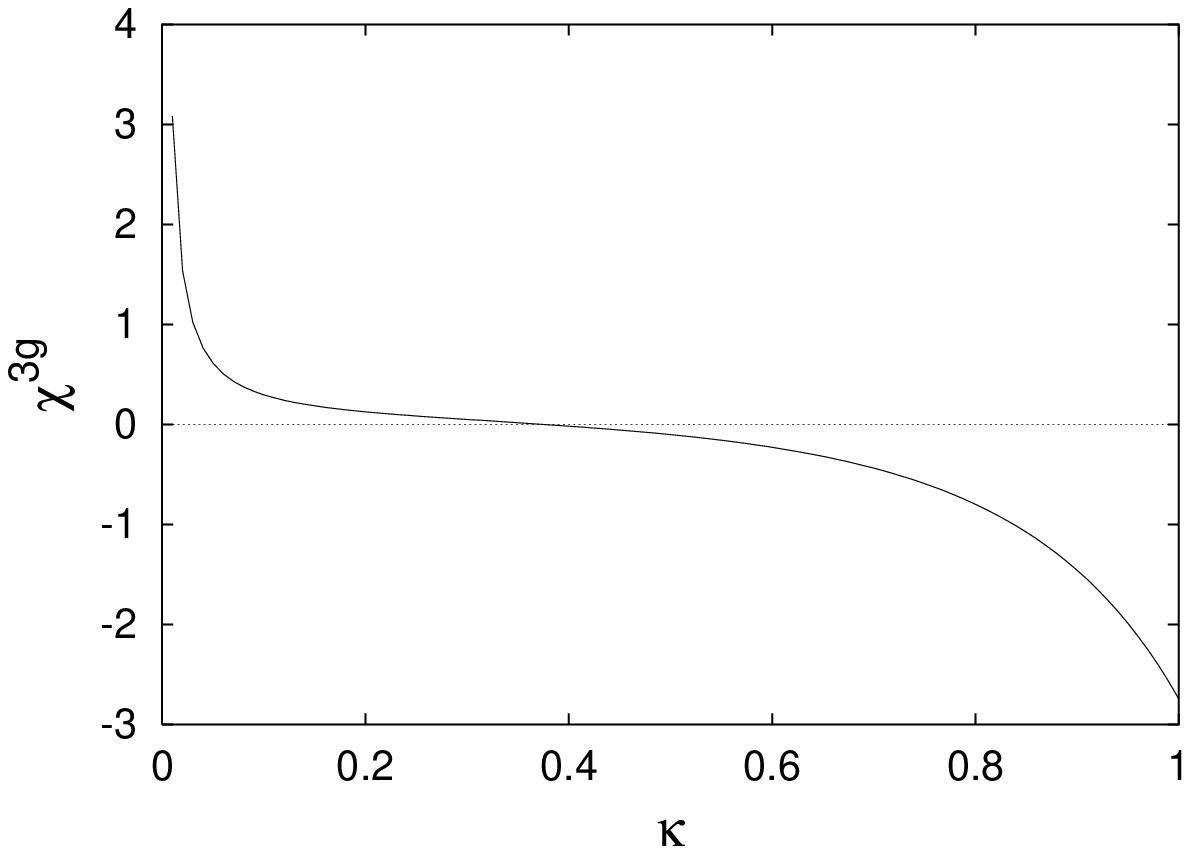}
\includegraphics[width=6.5cm]{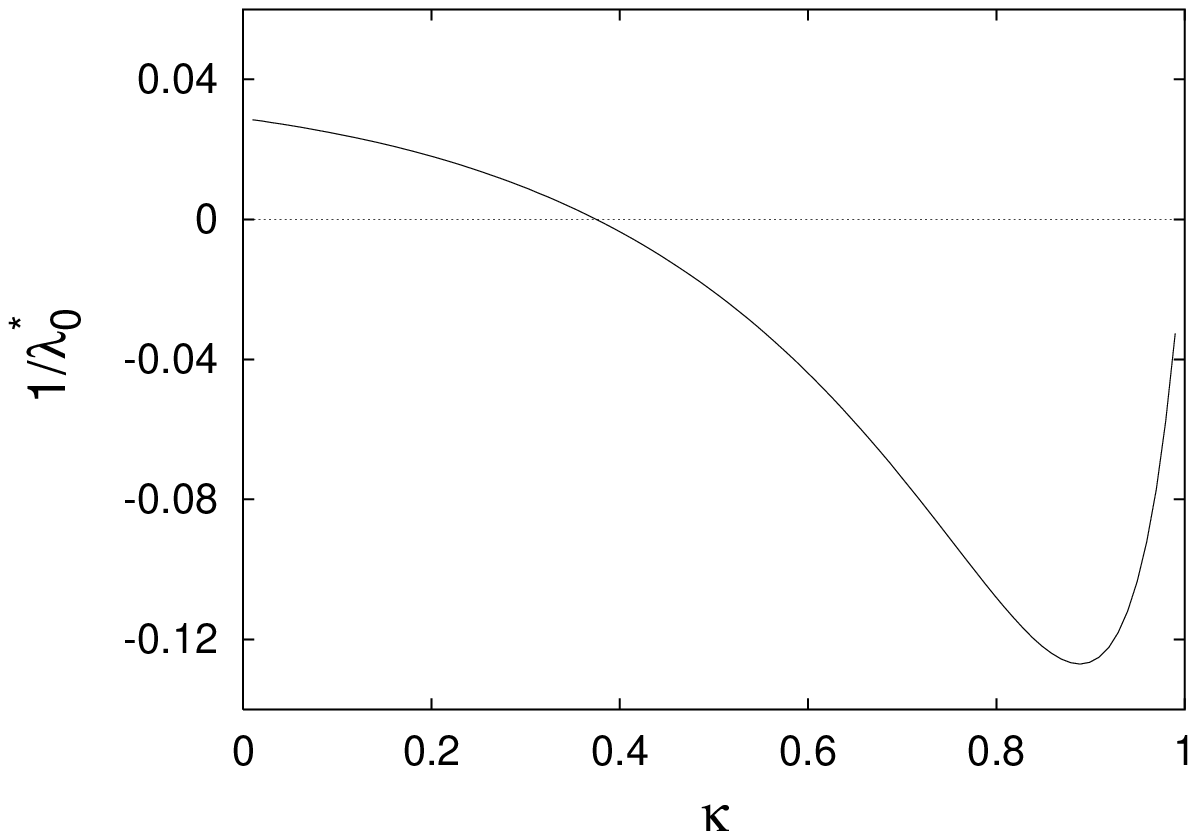}
\vspace{-2pt}
\caption{\label{Fig:chi3}
\textit{Left pane:} variation of the sunset contribution $\chiF^{3g}$ \eqref{chi3gb} as function of the infrared exponent $\kappa$.
\textit{Right pane:} Inverse of the infrared fixed point $\lambda_0^*$ \eqref{chiGl2} as function of the infrared exponent $\kappa$ without squint contribution.}
\end{figure}

First, the situation is investigated where all diagrams are taken into account \textit{except} the squint diagram.
The evolution of $\chiF^{3g}$ \eqref{chi3gb} as function of $\kappa$ is shown in the left pane of Fig.\ \ref{Fig:chi3}, and for example $\chiF^{3g}(0.5) \approx -0.1$. 
Table \ref{Table1} summarizes the various infrared contributions in Eq.\ \eqref{consist2}, with exception of $\chi^{4g}$, for $\kappa=0.25, 0.5, 0.75$, and in the right pane of Fig.\ \ref{Fig:chi3} the evolution of $1/\lambda_0^*$ from the gluon equation condition \eqref{chiGl2} is shown. The *-superscript is used to emphasize that the squint diagram was neglected. Clearly, no intersection will exist between the gluon curve of Fig.\ \ref{Fig:chi3} and any of the ghost curves of Fig.\ \ref{a0ghost}, and no power solution exists which simultaneously satisfies both ghost and gluon equations. 

\begin{table}[htb]
\beforetab
\begin{tabular}{l l r r r}
\firsthline
&& $\kappa = 0.25$ & $\kappa = 0.5$ & $\kappa = 0.75$ \\
\midhline
gluon & $\chiF^{2c}(\kappa)$ & $0.55$ & $0.5$ & $0.65$\\[1pt]
& $\chiF^{2g}(\kappa)$ & $-6.6$ & $-5.4$  & $-7.3$ \\[1pt]
& $\chiF^{3g}(\kappa)$ & $0.08$ & $-0.1$ & $-0.60$ \\[1pt]
& $1/\lambda_0^*(\kappa)$ & $0.01$ & $-0.02$ & $-0.09$ \\
\midhline
ghost & $\chiG^0(\kappa)$ & $2.92$ & $1.6$   & $1.21$ \\[1pt]
& $\chiG^\alpha(\kappa)$ &  $2.67$ & $1.2$  & $0.73$ \\
\lasthline
\end{tabular}
\aftertab
\captionaftertab[]{\label{Table1}Infrared contributions from the gluon vacuum polarization diagrams and ghost self-energy. An infrared solution requires $1/\lambda_0^*$ from the gluon equation to equate $\chi_G$ from the ghost equation.}
\end{table}

\begin{figure}[b]
\includegraphics[width=6.5cm]{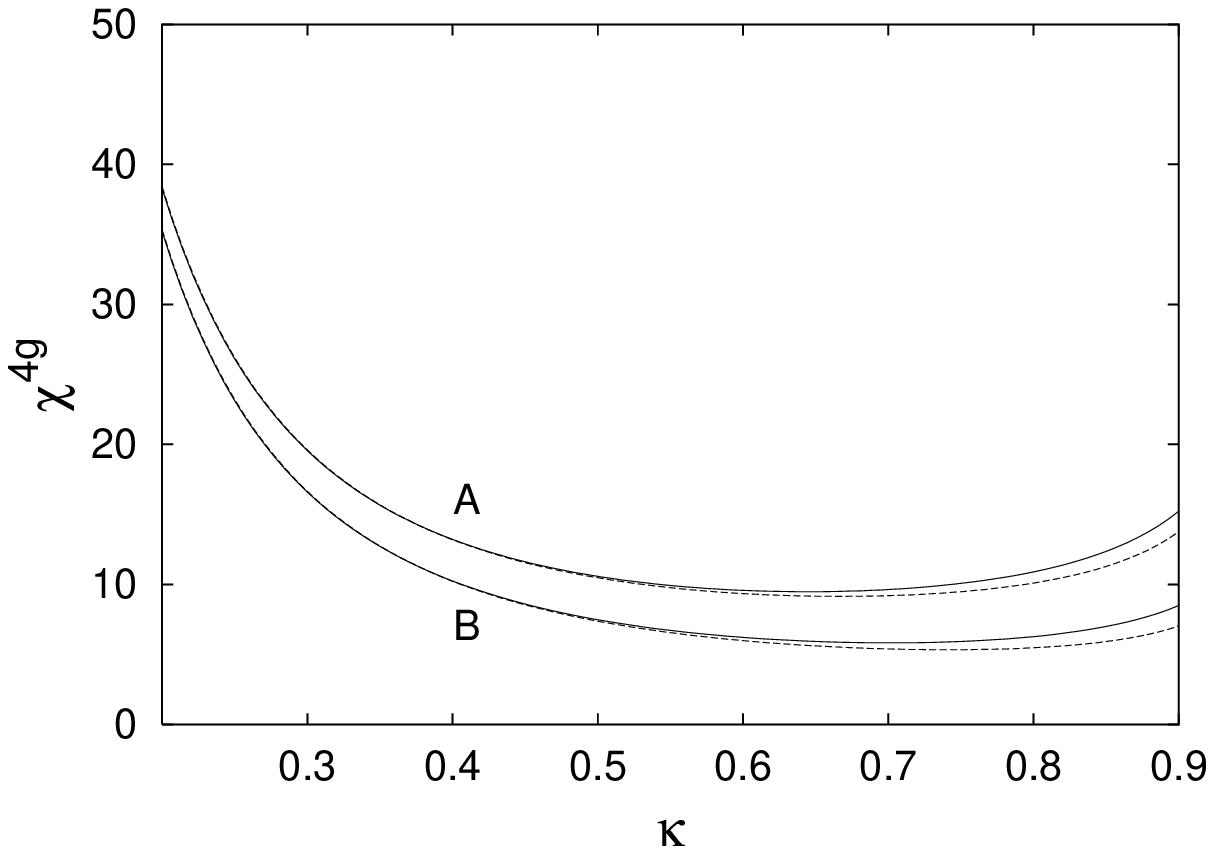}
\includegraphics[width=6.5cm]{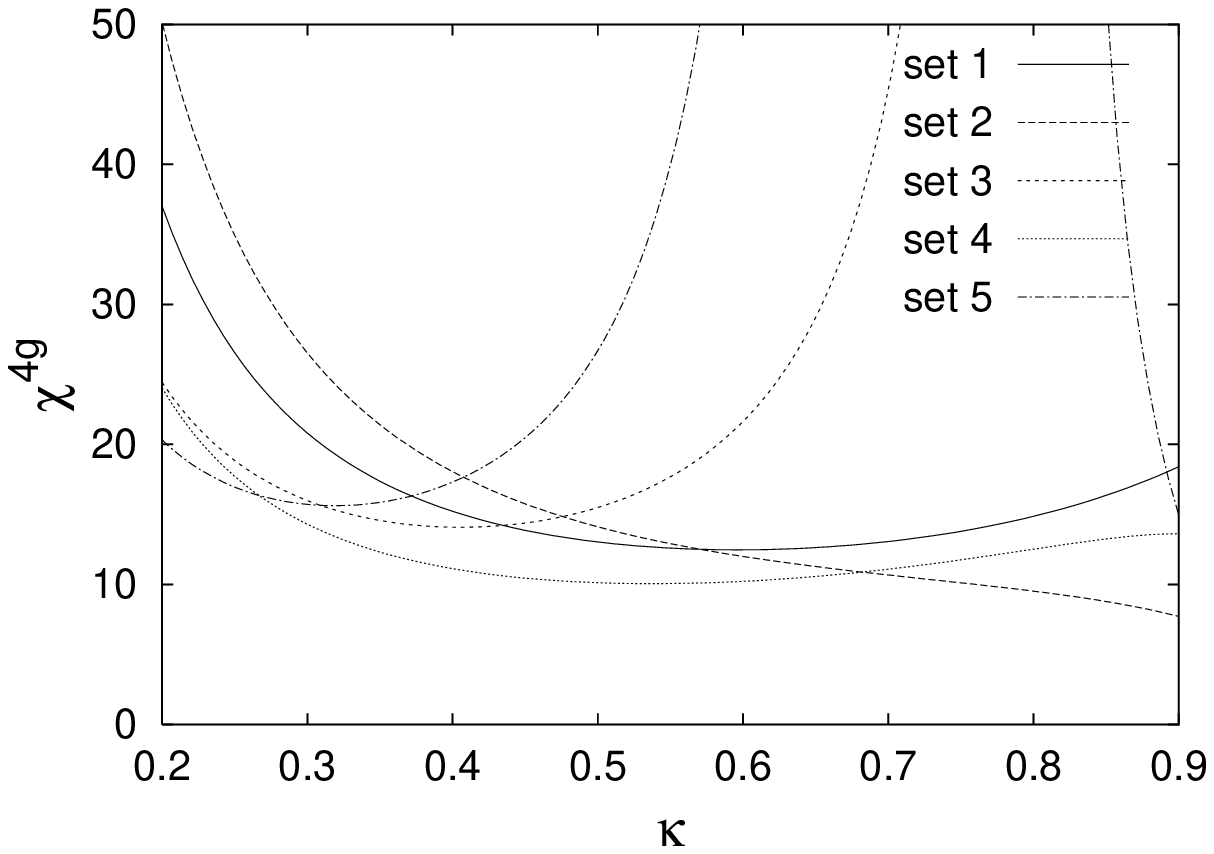}
\vspace{-2pt}
\caption{\label{chi4}\textit{Left pane:} Required value $\chiF^{4g}$ \eqref{chi4cond} as function of the infrared exponent $\kappa$ to ensure consistent infrared power solutions: (A) using the bare ghost equation truncation $\chiG^0$ \eqref{chi0}, (B) for the $\alpha$-truncation $\chiG^\alpha$ \eqref{chi-a}. The dashed curves, calculated without the sunset contribution, show that its contribution is negligible.
\label{chi4b}\textit{Right pane:} Squint contribution $\chiF^{4g}$ \eqref{chi4g} as function of the infrared exponent $\kappa$ for the parametrizations of Table \ref{Table2}.}
\end{figure}

Within the scope of the MR truncation, the only possible way left to find a consistent power solution to both equations is for the squint diagram to give the contribution necessary to balance both sides of Eq.\ \eqref{consist2}. A consistent infrared solution requires the squint diagram contribution to satisfy
\be\label{chi4cond}
\chiF^{4g}(\kappa) = \chiG^2(\kappa) - \chiG(\kappa)[\chiF^{2c}(\kappa)+\chiF^{2g}(\kappa)] - \chiF^{3g}(\kappa) \,,
\ee
and its variation with $\kappa$ is shown in the left pane of Fig.\ \ref{chi4}. The figure shows that the squint contribution is quite constrained for a solution to exist: it has to be positive and larger than 5. The figure also illustrates that the sunset diagram is negligible in the analysis of the infrared asymptotic behavior. 

As discussed in Sect.\ \ref{Sect:Truncation}, the MR truncation assumes the cancellation of a certain number of dressing function corrections in the loop kernels. The parameters $\mathbf a_{4g}$, $\mathbf b_{4g}$ of Eq.\ \eqref{4glooptrunc} determine how these corrections are distributed over the internal legs of the squint diagram.
Five simple choices are shown in Table \ref{Table2} as illustrations.
Set 1 is a simple equipartition of the dressing corrections over the various legs. Sets 2-5 correspond to attaching a $G/F$-correction to each three-gluon vertex and $G^2/F$ to the four-gluon vertex. The sets differ in the way the dressing corrections are distributed over the vertex legs. 

\begin{table}[ht]
\beforetab
\begin{tabular}{c l l l}
\firsthline
set & \multicolumn{1}{c}{$\mathbf a_{4g}$} & \multicolumn{1}{c}{$\mathbf b_{4g}$}  &   \multicolumn{1}{c}{$\mathbf k = \kappa(2 \mathbf b_{4g} - \mathbf a_{4g})$} \\
\midhline
1 & $(1,1,1,1)$ & $(1/4,1/4,1/4,1/4)$ & $(-\kappa/2,-\kappa/2,-\kappa/2,-\kappa/2)$  \\
2 & $(5/6,1,1,7/6)$ & $(1/6,1/3,1/3,1/6)$ & $(-\kappa/2,-\kappa/3,-\kappa/3,-5\kappa/6)$ \\
3 & $(5/6,4/3,4/3,1/2)$ & $(1/6,1/6,1/6,1/2)$ & $(-\kappa/2,-\kappa, -\kappa, \kappa/2)$ \\
4 & $(1/2,7/6,7/6,7/6)$ & $(1/2,1/6,1/6,1/6)$ & $(\kappa/2,-5\kappa/6,-5\kappa/6,-5\kappa/6)$ \\
5 & $(1/2,3/2,3/2,1/2)$ & $(1/2,0,0,1/2)$ & $(\kappa/2, -3\kappa/2,-3\kappa/2,\kappa/2)$ \\
\lasthline
\end{tabular}
\aftertab
\captionaftertab[]{\label{Table2}Various truncations of the squint diagram.}
\end{table}

\begin{figure}[b]
\begin{center}
\includegraphics[width=11.0cm]{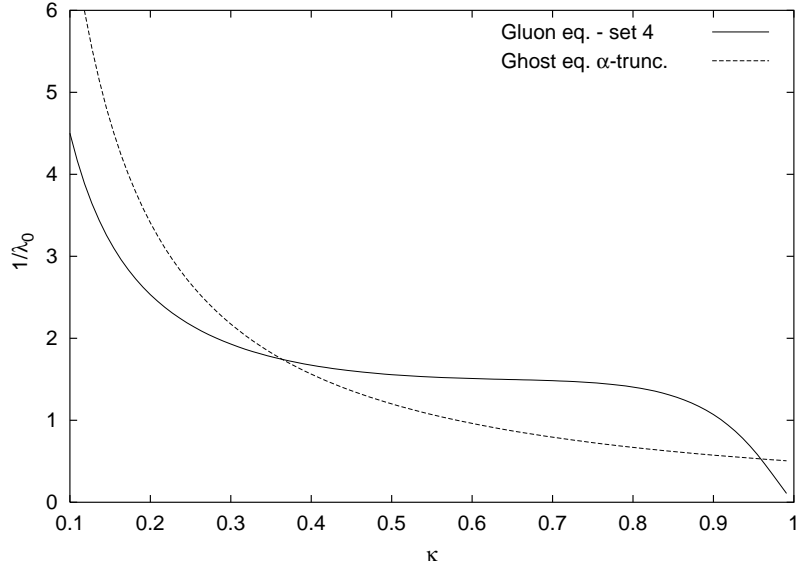}
\end{center}
\vspace{-15pt}
\caption{\label{chi4tot}Inverse fixed point $1/\lambda_0$ from the gluon equation, with squint parametrization set 4 of Table \ref{Table2}, and from the ghost equation in $\alpha$-truncation, as function of the infrared exponent $\kappa$. The intersection between gluon and ghost curves gives the solutions of the infrared consistency conditions. Two solutions are found: $\kappa$ = 0.36 and 0.96. Only the first one corresponds to a true solution of the integral equations, as was established from the numerical study.}
\end{figure}

As can be seen from Eq.\ (\ref{chi4g}) the \textit{infrared} contribution of the squint diagram depends on $\mathbf k = \kappa(2 \mathbf b_{4g} - \mathbf a_{4g})$, rather than on $\mathbf a_{4g}, \mathbf b_{4g}$ themselves, and the $k$-partitionings corresponding to the various parametrizations are also given in Table \ref{Table2}.
The infrared contributions $\chi^{4g}$ for these squint parametrizations are shown in the right pane of Fig.\ \ref{chi4b}. Comparing the required and actual squint contributions in the left and right panes of Fig.\ \ref{chi4} demonstrates that the squint diagram potentially has the right sign and magnitude to satisfy the infrared consistency conditions. Therefore, the value of $\kappa$ and $\lambda_0$ can be determined by looking for the intersection between the ghost and gluon curves. This is illustrated in Fig.\ \ref{chi4tot} for one specific squint truncation, i.e. set 4 of Table \ref{Table2}. 
Mathematica was used to determine the solutions of various truncations of the ghost-gluon system, constructed by combining all the squint truncations of Table \ref{Table2} with two different truncations of the ghost self-energy. Each truncation has up to two infrared solutions which are tabulated in Table \ref{Table3}. The numerical analysis of the coupled integral equations, which will be discussed in more detail later on, showed that not all infrared solutions correspond to consistent overall solutions connecting the infrared and the perturbative ultraviolet behaviors. The solutions with $\kappa \gtrapprox 0.75$ are spurious solutions of the infrared analysis, and only those with $\kappa \in [0.17,0.53]$  correspond to consistent solutions of the integral equations.

\begin{table}[t]
\beforetab
\begin{tabular}{c c c c c c c c}
\firsthline
   \multicolumn{2}{c}{truncation} & \multicolumn{3}{c}{solution 1} & \multicolumn{3}{c}{solution 2} \\
ghost & gluon & $\kappa$ & $\lambda_0$ &  numerical & $\kappa$ & $\lambda_0$ &  numerical \\
\midhline
A & $1$ & 0.24 & 0.32 & yes & 0.95 & 0.96 & no \\
& $2$ & -- & -- & -- & 0.78 & 0.84 & no \\
& $3$ & 0.38 & 0.50 & yes & --& -- & -- \\
& $4$ & 0.53 & 0.65 & yes & 0.90 & 0.92 & no \\
& $5$ & 0.35 & 0.46 & yes & 0.90 & 0.93 & no \\
\midhline
B & $1$ & 0.17  & 0.25 & yes & 0.98 & 1.95 & no \\
 & $2$ & -- & -- & -- & 0.91 & 1.77 & no \\
 & $3$ & 0.31 & 0.48 & yes & -- & -- & -- \\
 & $4$ & 0.36 & 0.57 & yes & 0.96 & 1.89 & no \\
 & $5$ & 0.31 & 0.48 & yes & 0.93 & 1.81 & no \\
\lasthline
\end{tabular}
\aftertab
\captionaftertab[]{\label{Table3}Solutions of the infrared consistency condition \eqref{consist2} and existence of numerical solutions for various truncations of the ghost-gluon system. Part A for the bare ghost self-energy truncation and part B for the $\alpha$-truncation. The squint truncations 1-5 are defined in Table \ref{Table2}.}
\end{table}

Clearly, the simple MR truncation is not able to pin down the exact QCD result: the value of $\kappa$ depends on the truncation and its range is quite broad. Nevertheless, the study elucidates the conspiracy between the various diagrams in the propagator equations, necessary to allow for consistent solutions of the coupled ghost-gluon system.

Now that power law solutions have been found satisfying the infrared consistency conditions, the next task is to solve the integral equations numerically and look for overall propagator solutions having such an infrared behavior. 
However, first a closer look is taken at the numerical treatment of the two-loop diagrams.

\section{Numerical Study}

\subsection{Approximated Two-loop Diagrams}
\label{Sect:EffLoop}

Although the infrared analysis of the two-loop diagrams could be performed analytically, their numerical treatment as part of an integral equation is a numerical problem of too high complexity to be accessed yet.
Therefore an alternative solution method is herein proposed where the two-loop integrals are replaced by effective one-loop integrals which have the same leading infrared behavior as the original integrals, are subleading in the perturbative region, and still satisfy the properties of multiplicative renormalizability. 

The MR truncated squint diagram was given in Eq.\ \eqref{4glooptrunc} by
\be\label{4gloopd}
\Pi^{4g}(x) = N_c^2 g_\mu^4  \intq{q_1} \intq{q_2} \, Q_4^0(p,q_1,q_2,q_3,q_4) \, G_R^4[\mathbf a_{4g}] \,  F_R^1[\mathbf b_{4g}] \,,
\ee
where $\mathbf a_{4g}=(a_1,a_2,a_3,a_4)$ and $\mathbf b_{4g}=(b_1,b_2,b_3,b_4)$ parametrize the squint truncation.

As discussed in the infrared analysis (see Sect. \ref{Sect:Two-loop}), the two-loop integration can be performed in two successive steps:
\be\label{4gloope}
\Pi^{4g}(x) = N_c^2 g_\mu^4  \intq{q_1} \Pi^{4g}_\text{inner}(x,z_1,z_4) \;  G_R^{a_1}(z_1) F_R^{b_1}(z_1)  G_R^{a_4}(z_4) F_R^{b_4}(z_4) \,,
\ee
with inner loop integral
\be\label{innerloop}
\Pi^{4g}_\text{inner}(x,z_1,z_4) = \intq{q_2} Q_4^0(p,q_1,q_2,q_3,q_4) \; G_R^{a_2}(z_2) F_R^{b_2}(z_2)  G_R^{a_3}(z_3) F_R^{b_3}(z_3) \,.
\ee

This two-step procedure provides an ideal opportunity to introduce effective one-loop integrals as approximation to the two-loop integrals, based on their analytic infrared evaluation. 
To evaluate the infrared contribution of the squint diagram, the dressing functions are replaced by their asymptotic power behaviors, and the inner loop \eqref{innerloop} can be written as
\be
\lim_{x\to0} \Pi^{4g}_\text{inner}(x,z_1,z_4) = A^{b_2+b_3} B^{a_2+a_3} \intq{q_2} Q_4^0(p,q_1,q_2,q_3,q_4) \; z_2^{k_2} z_3^{k_3} \,,
\ee
with $k_2=(2 b_2-a_2)\kappa$ and $k_3=(2 b_3-a_3)\kappa$. The integrals are evaluated as detailed in Sect. \ref{Sect:Two-loop}, and the solutions can be written as a function of squared external and outer loop momenta:
\be\label{Q4integ}
\lim_{x\to0} \Pi^{4g}_\text{inner}(x,z_1,z_4) = A^{b_2+b_3} B^{a_2+a_3} z_1^{k_2+k_3} \sum_{i,j,m} c_{i j m}(k_2,k_3) \, x^{i} z_1^{j} z_4^{m} \,,
\ee
where the coefficients $c_{i j m}$ are ratios of $\Gamma$-functions depending on $k_2, k_3$. 

The powers in front of the sum can be identified with the infrared behaviors of $G_R(z_1)$ and $F_R(z_1)$ and, in the infrared, the last expression is identical to
\be\label{pi-inner}
\lim_{x\to0} \Pi^{4g}_\text{inner}(x,z_1,z_4) = G_R^{a_2+a_3}(z_1) F_R^{b_2+b_3}(z_1) \sum_{i,j,m} c_{i j m}(k_2,k_3) \, x^{i} z_1^{j} z_4^{m} \,.
\ee

Although Eq.\ \eqref{pi-inner} is only exact in the infrared limit, it will be used to approximate the inner-loop integral \eqref{innerloop} for \textit{arbitrary} external momentum $x$, in order to construct an effective one-loop diagram replacing the original two-loop diagram \eqref{4gloope} in the numerical calculations.
Substituting Eq.\ \eqref{pi-inner} in Eq.\ \eqref{4gloope} yields the following effective squint contribution:
\be\label{eff2loop}
\Pi^{4g}_{\text{eff}}(x) \equiv N_c^2 g_\mu^4 \intq{q}  Q_4^{\text{eff}}(x,y,z) \; G_R^{a_y}(y) \, \, F_R^{b_y}(y) \,  G_R^{a_z}(z) \, F_R^{b_z}(z)  \,,
\ee
with
\be\label{ayby}
\begin{aligned}
a_y &= a_{1} + a_{2} + a_{3} \,, \\
b_y &= b_{1} + b_{2} + b_{3} \,, 
\end{aligned}\quad\qquad
\begin{aligned}
a_z &= a_{4}  \,, \\
b_z &= b_{4} \,,
\end{aligned}\hspace{1cm}
\ee
and where the effective kernel 
\be\label{Q4eff}
Q_4^{\text{eff}}(x,y,z) \equiv \sum c_{i j m}(k_2,k_3) \, x^{i} y^{j} z^{m} \,,
\ee
for arbitrary external momentum $x$, is defined by the infrared integral of the inner loop, given in Eq.\ \eqref{Q4integ}. Its explicit form is given in \ref{App:Eff2loop}. 

Because of the properties of $\mathbf a_{4g}$ and $\mathbf b_{4g}$, the exponents in the effective loop obey
\be\label{rFy}
a_y + a_z = 4  \quad \, \qquad b_y + b_z = 1 \,,
\ee
so that the approximation conserves the number of dressing functions of the original integral. This is sufficient to ensure that the approximation remains multiplicatively renormalizable and ultraviolet subleading. Furthermore, the exponents also satisfy 
\be\label{rFysquint}
(2b_y - a_y)\kappa = k_1+k_2+k_3 \quad , \qquad
(2b_z - a_z)\kappa = k_4 \,,
\ee
and the effective loop reproduces the same infrared result \eqref{chi4g} as the original two-loop diagram. The preservation of these properties is obvious by construction.
 
Away from the infrared limit, the propagators no longer behave as power laws, and therefore the infrared powers in Eq.\ \eqref{Q4integ} were replaced by the corresponding dressing functions in Eq.\ \eqref{pi-inner} when devising the approximation. This guarantees that the ultraviolet behavior is logarithmically subleading and avoids contamination with unjustifiable power behaviors.
Indeed, substituting the leading-order perturbative results $F(x) \sim \alpha(x)^{-\gamma}$ and $G(x) \sim \alpha(x)^{-\delta}$, with $\gamma=-13/22$, $\delta=-9/44$, in the approximated two-loop diagram \eqref{eff2loop} yields an ultraviolet contribution of order $\alpha(x)^{1-\gamma}$, which has the correct log-exponent of the resummed next-to-leading order correction to the gluon anomalous dimension \cite{Altarelli:1982ax}. Hence, the effective loop correctly mimics the ultraviolet behavior of the two-loop diagram, albeit not with the correct coefficient.

Although Eq.\ (\ref{eff2loop}) looks similar to the ghost and gluon loop contributions (\ref{pightrunc}) and (\ref{pigltrunc}), there is a {\it crucial} difference between the genuine one-loop integrals and the effective loop diagram. The latter contains an additional factor $N_c \alpha_\mu F_R(...) G_R^2(...)$ which behaves as a running coupling $\alpha(y)$ in the ultraviolet and ensures that the loop is logarithmically subleading. 

The effective sunset contribution is constructed in exactly the same way as depicted above for the squint diagram.
Note that the coefficients $c_{i j m}$ of the effective kernels for the sunset diagram are an order of magnitude smaller than those of the squint diagram, which agrees with the earlier conclusion that its infrared contribution is negligible.  

The introduction of the effective loops allows for a numerical study of the coupled ghost-gluon system including the main features of the two-loop diagrams to the gluon vacuum polarization.

\subsection{Results}

The existence of power laws satisfying the infrared consistency conditions does not yet guarantee that the coupled integral equations have solutions over the whole momentum region connecting the predicted infrared asymptotic behavior with the ultraviolet perturbative regime. A straightforward way to answer this question is to look numerically for solutions of the coupled equations. 

To solve Eqs.\ \eqref{SD-Gh} and \eqref{SD-Gl} numerically the unknown seeds $\tilde Z_3$ and $Z_3$ of the equations are eliminated by subtracting each equation at a fixed subtraction point:
\be
\label{subtracted}
\begin{aligned}
\frac{1}{F_R(p^2)} &= \frac{1}{F_R(\sigma_F^2)} - \Pi(p^2) + \Pi(\sigma_F^2)\,, \\
\frac{1}{G_R(p^2)} &= \frac{1}{G_R(\sigma_G^2)}  - \Sigma(p^2) +  \Sigma(\sigma_G^2) \,,
\end{aligned}
\ee
and the renormalization conditions are fixed\footnote{For reasons of numerical stability, the ghost equation is subtracted at $\sigma_G^2=0$, where $1/G_R(0)=0$, and the renormalization conditions are chosen by fixing $\alpha_\mu$, $F_R(\sigma^2)=F_\sigma$, and the coefficient $A$ of the asymptotic infrared power law for $F(x)$. Because of multiplicative renormalizability these conditions are equivalent to the conditions $\alpha(\mu^2)=\alpha_\mu, G_R(\mu^2)=F_R(\mu^2)=1$.} by requiring $\alpha(\mu^2)=\alpha_\mu$ and  $G_R(\mu^2)=F_R(\mu^2)=1$.

\begin{figure}[htp]
\includegraphics[width=6.5cm]{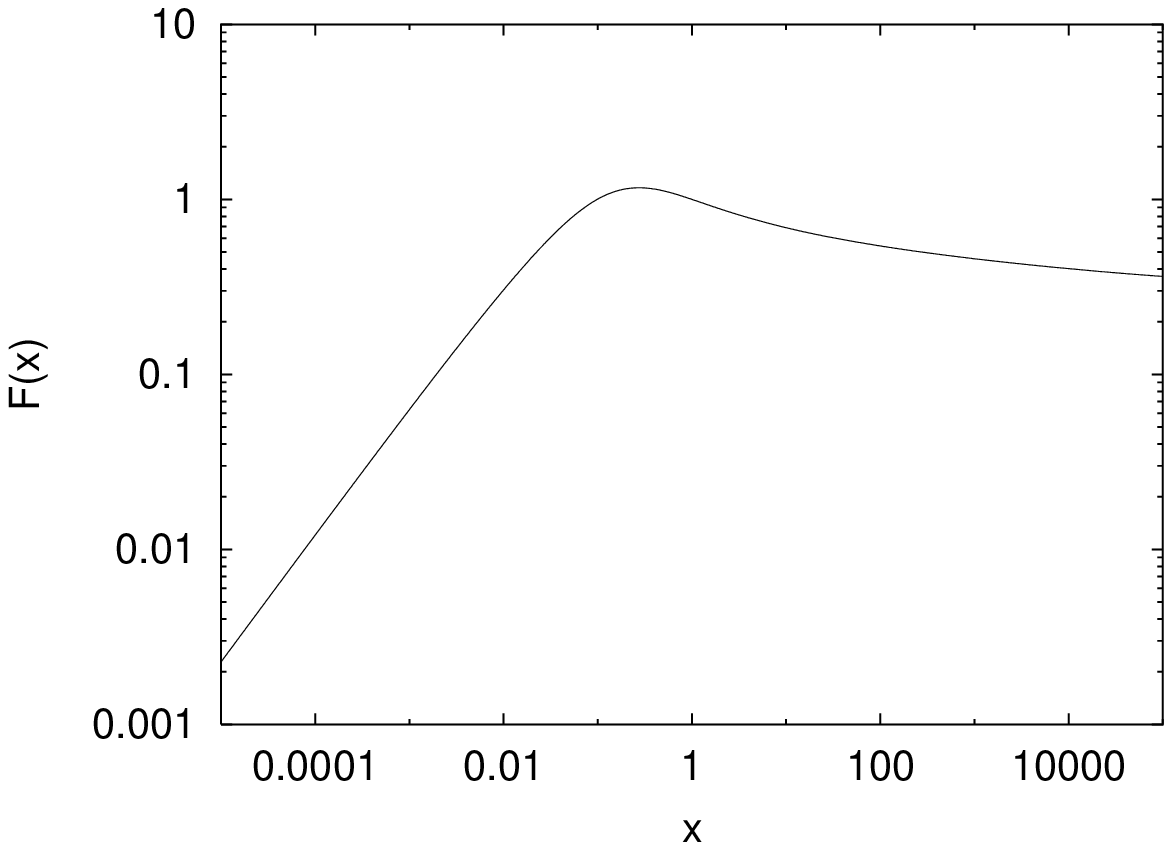}
\includegraphics[width=6.5cm]{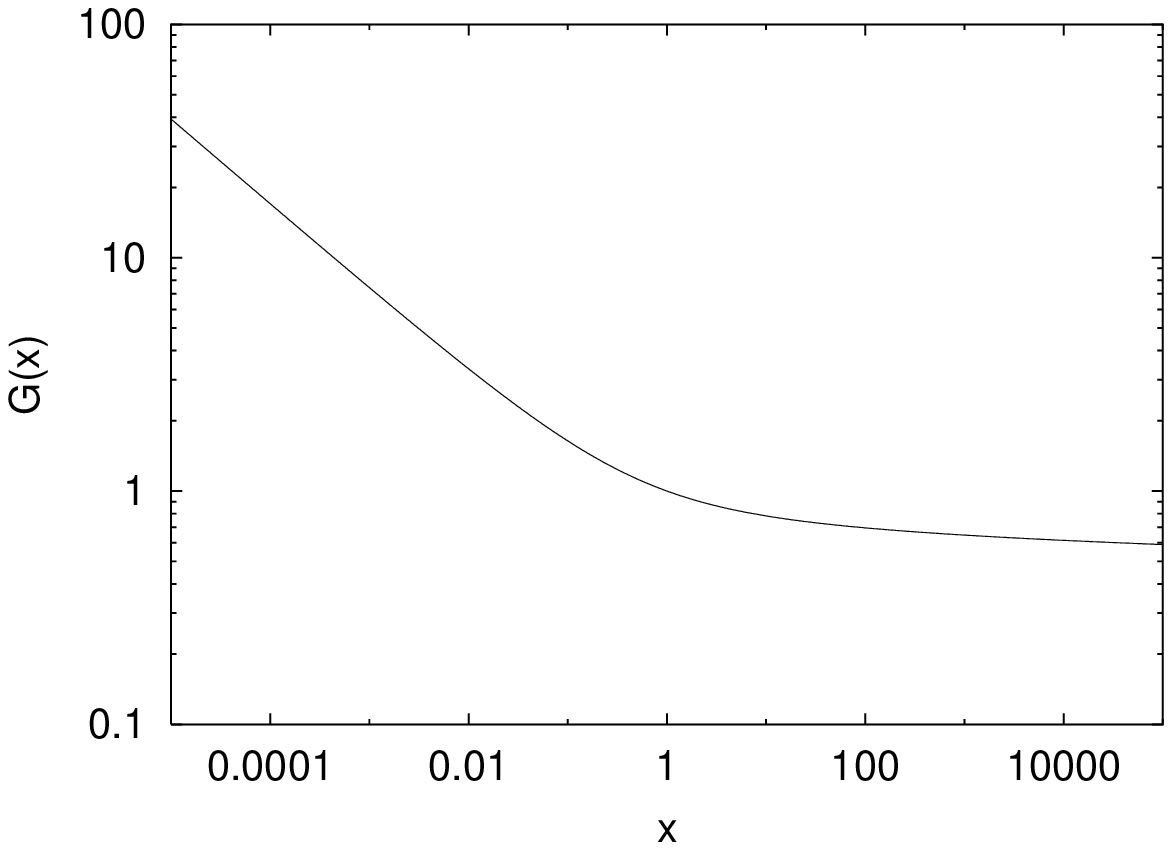}
\vspace{-4pt}
\caption{Numerical solutions for truncation B.4 of Table \ref{Table3} with $\kappa=0.36$. \label{Fig:F}\textit{Left pane:} Gluon dressing functions as function of squared momentum, with renormalization condition $F(1)=1$.
\label{Fig:G}\textit{Right pane:} Ghost dressing functions as function of squared momentum, with $G(1)=1$.}
\vspace{9pt}
\includegraphics[width=8.0cm]{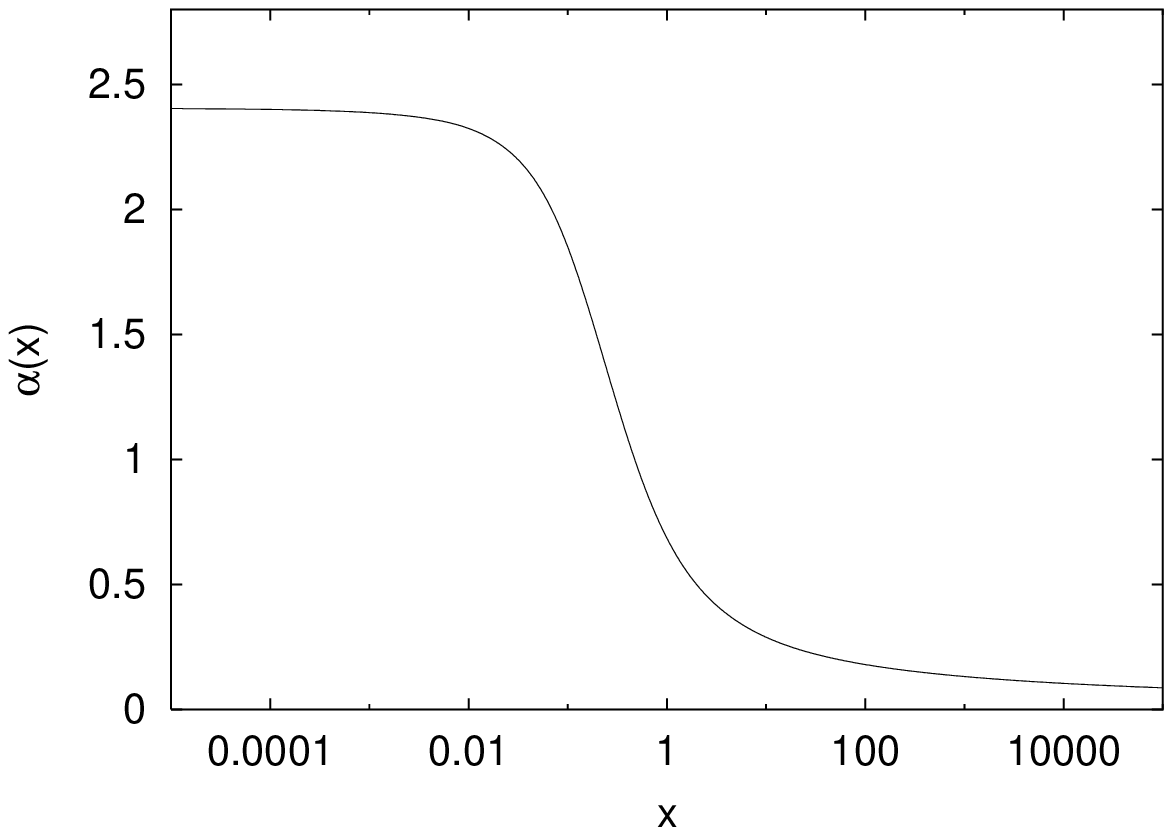}
\parbox[b]{5.3cm}{\caption{\label{Fig:alpha}Running coupling for truncation B.4 of Table \ref{Table3}, computed using Eq.\ \eqref{rgi2} and plotted as function of squared momentum. The infrared fixed point value is $\alpha_0=2.4$.}}
\end{figure}

The numerical method employed to solve the set of coupled integral equations \eqref{subtracted} was developed in Ref.\ \cite{Bloch:1995dd} and is described in more detail in \ref{App:NumMethod}. 
Numerical solutions were sought for the ten truncations of the ghost-gluon system described in Table \ref{Table3} of Sect.\ \ref{Sect:consist}. As shown in that table, the lowest lying $\kappa$ of each truncation corresponds to a consistent numerical solution, while the higher value is spurious as no numerical solution having such an infrared behavior is found. 
The numerical solutions for the various parametrizations are very similar to one another, and typical SU(3) solutions\footnote{The relation between SU(N) and SU(3) results is trivial as can be seen from the equations of Sect.\ \ref{Sect:Equations} and the discussion after Eq.\ \eqref{lambda}. The solutions for $F$ and $G$ remain unchanged, while the running coupling $\alpha$ is multiplied by a factor 3/N.} for truncation B.4 of Table \ref{Table3} are shown in Figs.\ \ref{Fig:F} and \ref{Fig:alpha}. 
The infrared power behaviors of the dressing functions and the infrared fixed point of the running coupling are in perfect agreement with the predictions of the infrared analysis, with $\kappa=0.36$ and $\alpha_0=2.4$ for SU(3). Note that the infrared fixed point of the running coupling in Fig.\ \ref{Fig:alpha} is the result of a product of power laws with vanishing total exponent (see Eq.\ \eqref{nu}), and its truthful rendering requires a numerical method with high accuracy.

Because the equations are scale invariant, the momentum units of the solutions are arbitrary, and the scale should be fixed using experimental input. A typical choice is to fix the coupling at a specific perturbative mass scale, for instance $\alpha(M_Z^2)$. However, as this is a study of pure gauge theory and no quark loops are taken into account, the coupling does not run like the physical one, and such a scale matching would have no phenomenological use. Therefore, the momentum units in Figs.\ \ref{Fig:F} and \ref{Fig:alpha} are left arbitrary for the time being. Nevertheless, it is clear that the solutions contain an intrinsic scale, which signals the transition from the perturbative to the nonperturbative region and is fixed by the renormalization conditions. Note that because the equations satisfy multiplicative renormalizability the solutions for \textit{arbitrary} renormalization conditions can easily be derived from the solutions computed for a particular renormalization. 

Of major interest is the comparison between recent SU(2) lattice simulations \cite{Bonnet:2000kw, Langfeld:2001cz, Bloch:2002we} and the Dyson-Schwinger results. The lattice data also exhibit infrared power behaviors for the gluon and ghost dressing functions, and they seem to indicate that $\kappa = 0.5$ as the gluon propagator goes to a non-zero, finite value at zero momentum. 
Furthermore, the lattice results feature a pronounced bump in the gluon dressing function, in the transition from the perturbative to the nonperturbative behavior. 

Because of its simplicity, the MR truncation cannot be expected to \textit{predict} the correct value of the infrared exponent $\kappa$. Moreover, for a given $\kappa$ the effective one-loop approximation to the two-loop diagrams (see Sect.\ \ref{Sect:EffLoop}) most probably affects the intermediate behavior of the solutions. An important goal of this investigation is to find out to what extent the two-loop improved MR truncation of the ghost-gluon system is able to accommodate the various features exhibited by the lattice data. For this purpose the physically motivated parametrizations of the squint diagram proposed in Table \ref{Table2} are now relaxed, and a $\kappa=0.5$ solution is constructed by tuning the parameters. Note that the numerical solutions of the Dyson-Schwinger equations are stable with respect to variations of the parameters. The main effects are changes in the infrared exponent and the related value of the infrared fixed point of the coupling, and changes in the steepness of the transition from perturbative to nonperturbative region, which is related to the height of the bump in the gluon dressing function.

The MR truncation requires the specification of the ghost self-energy truncation as discussed in Sect.\ \ref{sect:ghost}. As illustrated in Fig.\ \ref{a0ghost} the value of the infrared fixed point of the coupling is related to the value of $\kappa$ through the ghost equation conditions \eqref{chi0}-\eqref{chi-a}. For $\kappa=0.5$ the $\alpha$-truncation \eqref{chi-a} yields a higher value $\alpha_0 = 5.2$ for the SU(2) infrared fixed point (compared to $3.9$ for the bare truncation), and seems in better agreement with the lattice data \cite{Bloch:2002we}.

The squint diagram truncation is parametrized by the sets $\mathbf a_{4g}$ and $\mathbf b_{4g}$ of Eq.\ \eqref{4glooptrunc}, and its infrared contribution is determined by $\mathbf k = \kappa(2\mathbf b_{4g}-\mathbf a_{4g})$. In practice, a $\kappa=0.5$ solution is constructed by 
solving the consistency equation (\ref{consist2}) for $\mathbf k$, after imposing the conditions $k_1+k_2+k_3+k_4=-2\kappa$ and $k_2=k_3$ (symmetry of the inner loop). 
The solution is not unique and one possible solution, which will be investigated in more detail, is $\mathbf k$ = ($-2.072,0.536,0.536,0$). 

The effective one-loop integrals \eqref{eff2loop} approximating the two-loop integrals in the numerical calculation, depend on $k_2$ and $k_3$, which fix the coefficients $c_{ijm}$ of the effective integration kernel \eqref{Q4eff}, and on $a_y$, $a_z$, $b_y$ and $b_z$ defined in Eq.\ \eqref{ayby}. 
Given a set $\mathbf k$, determining the infrared behavior of the squint diagram, there is one more parameter from the set $a_y$, $a_z$, $b_y$, $b_z$ that can be chosen freely. The remaining parameters are then fixed by Eqs.\ \eqref{rFy} and \eqref{rFysquint}. This freedom of choice leaves the leading infrared and ultraviolet behaviors unchanged, but can influence the behavior of the solutions in the intermediate momentum region. Hence, it can be used as phenomenological parameter to recover some of the information that is unavoidably lost by replacing the two-loop integrals by effective integrals. 
In practice, the parameter $b_y$ is tuned to reproduce the height of the bump in the lattice gluon dressing function, and good agreement is found for $b_y=-0.6$, which yields $a_y=0.8$, $a_z=3.2$ and $b_z=1.6$. 

\begin{figure}[htb]
\begin{center}
\includegraphics[width=11cm]{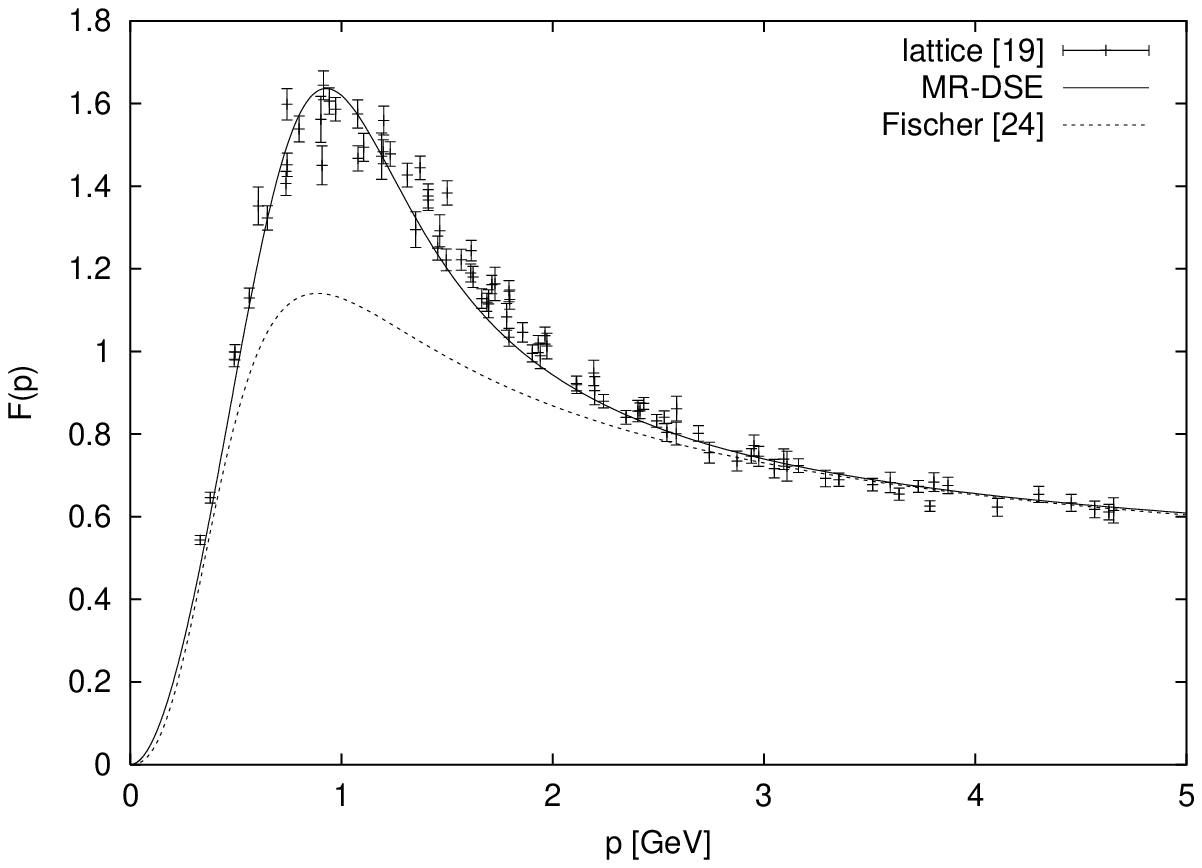}\vspace{10pt}
\includegraphics[width=11cm]{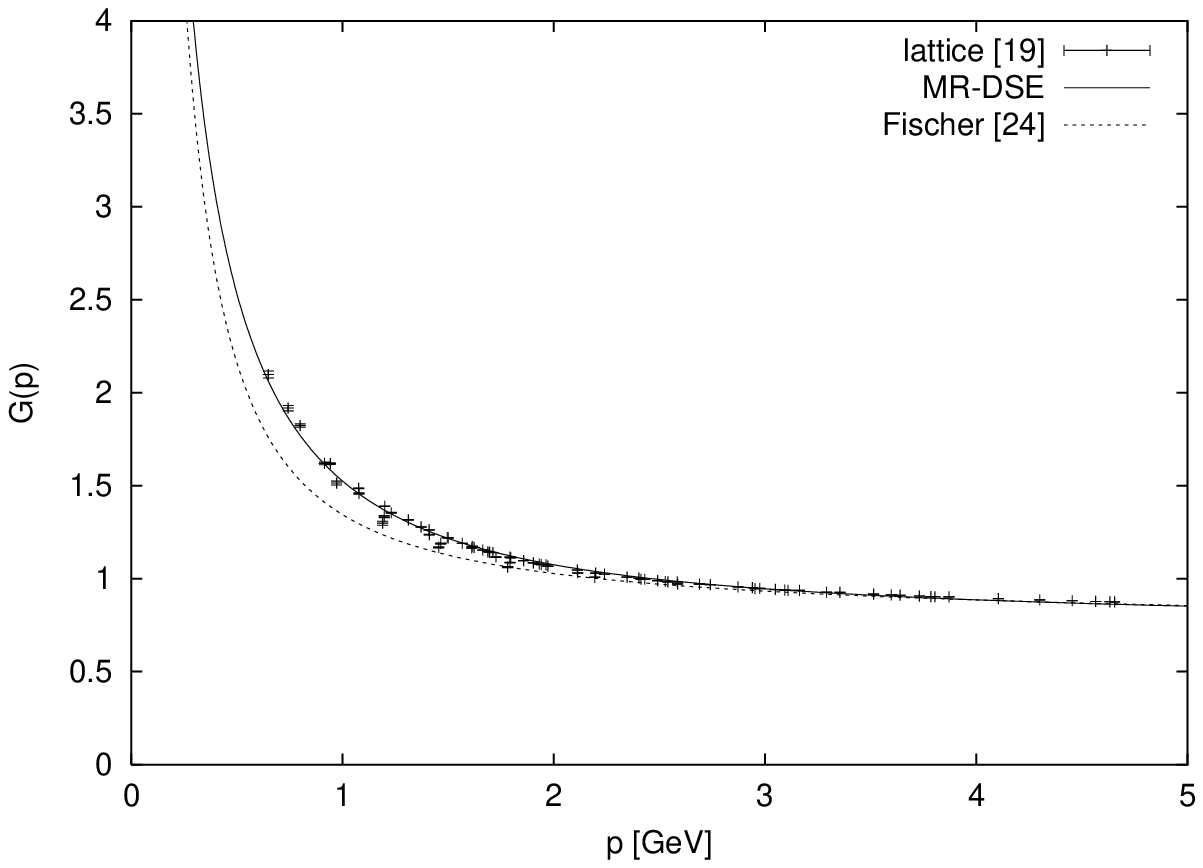}
\end{center}
\vspace{-15pt}
\caption{\label{Fig:Fcomp}\label{Fig:Gcomp}Comparison of the lattice data from ref.\ \cite{Bloch:2002we} with the DSE results using the two-loop improved MR truncation (with $\kappa=0.5$) and from ref.\ \cite{Fischer:2002hn}. \textit{Upper pane:} Gluon dressing functions as function of momentum. \textit{Lower pane:} Ghost dressing functions as function of momentum. }
\end{figure}

\begin{figure}[tb]
\begin{center}
\includegraphics[width=11cm]{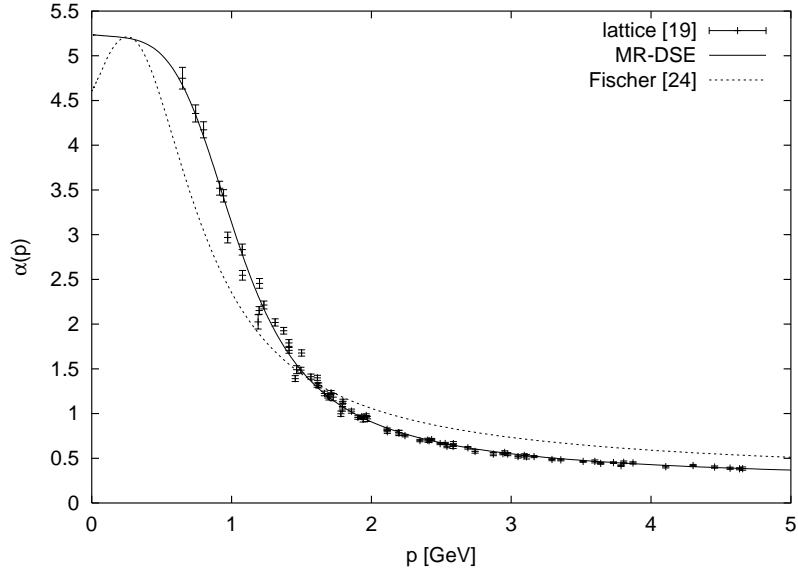}
\end{center}
\vspace{-15pt}
\caption{\label{Fig:alphacomp}Running coupling as function of momentum for SU(2). Comparison of the lattice data from ref.\ \cite{Bloch:2002we} with the DSE results using the two-loop improved MR truncation (with $\kappa=0.5$) and from ref.\ \cite{Fischer:2002hn}. The fixed point of the MR truncation is $\alpha_0=5.24$.}
\end{figure}

The DSE results are compared with the SU(2) lattice data of ref.\ \cite{Bloch:2002we}. These lattice simulations were performed on a $16^3 \times 32$ grid using 6 different $\beta$-values ranging from 2.1 to 2.5. The momentum scale of the lattice calculations is fixed using a string tension of $440 \, \MeV$. The momentum units of the DSE results are fixed by maximizing the overlap between the DSE solution for the running coupling and the corresponding lattice data. The running coupling on the lattice was also computed from the dressing functions using Eq.\ \eqref{rgi2} (see ref.\ \cite{Bloch:2002we}).
Figures \ref{Fig:Fcomp} and \ref{Fig:alphacomp} compare the lattice data with the DSE results\footnote{The data are now plotted on linear scale for both axes as the lattice data cover a much smaller momentum region than the DSE computations. Also note that the abscissa is $p$ instead of $p^2$.} computed with the two-loop improved MR truncation, and with the truncation of ref.\ \cite{Fischer:2002hn}.   

The ability for the DSE formalism to reproduce the lattice data, by introducing the two-loop improved MR truncation, is way beyond the initial expectations. 
Even though the DSE truncation was fixed using some characteristics of the lattice results,
the DSE solutions are not molded to those on the lattice, and they genuinely predict a number of additional features which are consistent with the lattice data, as is discussed below.

Although the height of the bump in the gluon dressing function  (upper pane of Fig.\ \ref{Fig:Fcomp}) was tuned to that of the lattice, the position and width of the peak are genuine DSE predictions which agree well with the lattice data. The ghost dressing function (lower pane of Fig.\ \ref{Fig:Gcomp}) is in total agreement with the lattice data, and reproduces well the smooth evolution from the perturbative region into the $1/p$ infrared singularity.

Even more remarkable is the agreement of the running coupling (Fig.\ \ref{Fig:alphacomp}) over the complete momentum region.
The value of the fixed point is a DSE prediction, determined by the ghost infrared condition \eqref{chiGh}.
Even though the lattice simulations are not yet sampling the infrared region deep enough to clearly show the bend-over towards a fixed point, the existence and value of the DSE fixed point are consistent with the actual infrared behavior of the lattice data.  
In the DSE study, the steepness of $\alpha(x)$ in the transition region appears to be tightly related to the height of the bump in the gluon dressing. 
The computations show that the steepness of the DSE coupling agrees well with the lattice data precisely when the height of the gluon bump does too, and hence, their interrelation is correctly reproduced by the DSE truncation.
The perturbative tail agrees very well with the lattice results too, illustrating that the solutions of the two-loop improved DSE truncation go beyond the leading logarithm. This contrasts with the usual one-loop inspired truncations, which typically yield ultraviolet tails that are 20\%
larger than the lattice results, as can be seen in Fig.\ \ref{Fig:alphacomp} from the DSE curve taken from ref.\ \cite{Fischer:2002hn}. 

It is also useful to construct a fit to the running coupling solution, which could be used in hadronic phenomenology:
\be\label{alphafit}
\alpha(x) = \frac{1}{c_0+x^2}\left[(c_0 + c_1 x^{\rho}) \, \alpha_0 +  x^2 \, \alpha_{2\ell}(x+\tau) \right] \,,
\ee
with $x=p^2/\LQCDsq$, and $\alpha_{2\ell}(x)$ is the two-loop perturbative running coupling \cite{Hinchliffe:2002ex}. The fixed point $\alpha_0=5.24$ is known from the infrared analysis, $\tau$ is set equal to 2, and the other parameters are fitted to the DSE solution of Fig.\ \ref{Fig:alphacomp}: $\LQCD=856 \, \MeV$, $c_0=1.16$, $\rho=0.66$ and $c_1=-0.070$. 

\begin{figure}[ht]
\includegraphics[width=7.7cm]{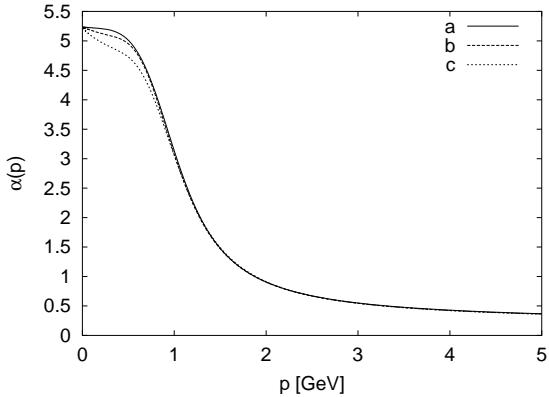}
\parbox[b]{5.6cm}{
\caption{\label{Fig:alphaDSE}Running coupling for three different parametrizations, constructed to yield $\kappa=0.5$ and reproduce the bump in the gluon dressing function, and specified in the table below.}
\scriptsize
\begin{tabular}[b]{l l l}
\firsthline
 & \multicolumn{1}{c}{$\mathbf k$} & \multicolumn{1}{c}{$b_y$} \\
\midhline
a & ($-2.072,0.536,0.536,0$) & -0.6 \\
b & ($-1.552, 0.526, 0.526, -0.5$) & -0.15 \\
c & ($0.596,-0.5,-0.5,-0.596$)  & -0.1\\
\lasthline
\end{tabular}
\normalsize
}
\end{figure}

Note that other truncation parametrizations, which also give $\kappa=0.5$ and reproduce the gluon bump, yield solutions which only differ in the nonperturbative approach to the fixed point of the running coupling, as shown in Fig.\ \ref{Fig:alphaDSE}. 

The results presented in Figs.\ \ref{Fig:Fcomp} and \ref{Fig:alphacomp} are solutions for pure SU(2) gauge theory. 
For nature's QCD the corresponding SU(3) value of the fixed point is $\alpha_0=3.49$. 
Furthermore, to get the correct running of the coupling, quark loops have to be included in the gluon vacuum polarization taking into account their dynamically generated mass. This is done by solving the coupled quark-ghost-gluon system in an MR truncation merging the truncation presented herein with those for the quark loop \cite{Bloch:2001wz} and for the quark equation  \cite{Bloch:2002eq}.
The solution of this coupled system will be reported elsewhere \cite{Bloch:2002c}, but from a previous study of the quark equation \cite{Bloch:2002eq} it is already clear that a running coupling similar to Eq.\ \eqref{alphafit} is capable of producing a strong enough breaking of the chiral symmetry to allow for consistent hadronic phenomenology.

\section{Conclusion}

In ref.\ \cite{Bloch:2001wz} a truncation of the Landau gauge Dyson-Schwinger equations for the ghost and gluon propagators was proposed, which respects the principles of multiplicative renormalizability and reproduces the leading order resummed perturbative results. This was achieved by directly approximating the integration kernels, rather than introducing explicit approximations to the full vertices and their corresponding renormalization constants. A remarkable feature of the truncation is that all diagrams in the gluon vacuum polarization contribute to the leading power in the infrared analysis. However, in that study the contributions of the two-loop gluon vacuum polarization diagrams could not be evaluated, and the MR truncation did not yield solutions for the coupled ghost-gluon system when only taking into account the one-loop diagrams. 

The present paper generalized further the MR truncation scheme of ref.\ \cite{Bloch:2001wz} and confirmed that the interplay between all the loops in the gluon vacuum polarization is a key ingredient to find consistent MR solutions of the coupled ghost-gluon system.  

An analytic method was developed to compute the contributions of the two-loop diagrams to the leading infrared power behavior of the gluon vacuum polarization, such that a complete infrared analysis of the coupled ghost-gluon system could be performed. The investigation showed that the squint diagram is able to supply an infrared contribution that allows for propagator solutions with infrared power behavior. The gluon and ghost dressing functions respectively behave like $(p^2)^{2\kappa}$ and  $(p^2)^{-\kappa}$, and the running coupling exhibits an infrared fixed point. The value of the infrared exponent depends on the details of the truncation, and some simple realizations of the truncation yield values of $\kappa \in [0.17,0.53]$. 

The numerical study of the coupled ghost-gluon system was facilitated by approximating the two-loop gluon vacuum polarization by an effective one-loop integral with the same leading infrared behavior. The numerical analysis produced solutions of the ghost-gluon system over the whole momentum range, and their infrared behaviors agree with the predictions of the analytic infrared analysis. 

Recent lattice simulations suggest that $\kappa=0.5$, and DSE solutions having such an infrared behavior were constructed by adjusting the truncation parameters. The numerical solutions for the dressing functions and running coupling showed that the two-loop improved MR truncation of the Dyson-Schwinger equations is able to produce solutions that are in extremely good agreement with the lattice data. 

This paper presented a study of the pure gauge sector. In a forthcoming study \cite{Bloch:2002c}, this MR truncation will be merged with that of the quark equation developed in ref.\ \cite{Bloch:2002eq}, to allow for a complete self-consistent study of dynamical chiral symmetry breaking in QCD.

\begin{acknowledge}

I would like to thank R.\ Alkofer, D.\ Atkinson, C.\ Fischer, K.-I.\ Kondo, K.\ Langfeld, S.M.\ Schmidt and P.\ Watson for useful comments and discussions. Most of this work was done at the Institute for Theoretical Physics of T\"ubingen University, and funded by the Deutsche Forschungsgemeinschaft under project no. SCHM 1342/3-1. It was also partly supported by the DFG research center "Mathematics for Key
Technologies" (FZT 86) in Berlin.

\end{acknowledge}

\appendix
\makeatletter
\renewcommand\@seccntformat[1]{{\csname
the#1\endcsname}\protect:\hspace{0.5em}}
\renewcommand\thesection{Appendix \Alph{section}}
\makeatother

\section{Color Contraction}
\label{App:colorcontractions}

The color contraction identities needed in Sect.\ \ref{Sect:Equations} are
\begin{align}
& f_{c_1 c_2 a} f_{c_1 c_2 b} = C_A \delta_{a b} \,,\\
& f_{c_1 c_2 a} f_{c_1 c_3 b} f_{c_2 c_3 c} = \frac{C_A}{2} f_{a b c} \,,\\
& f_{a c_1 c_2} f_{b c_3 c_4} f_{c_1 c_3 c_5} f_{c_2 c_4 c_5} = \frac{C_A^2}{2} \delta_{a b} \,.
\end{align}

\section{Multiplicative Renormalizability in Quenched QED}
\label{App:QED}

To get some feeling for the truncation suggested in Sect.\ \ref{Sect:Truncation}, it is good to make a detour over quenched QED \cite{Curtis:1990zs} and see how there the multiplicative renormalizability of the massless quark propagator is achieved in the Dyson-Schwinger equation, at least on the perturbative level.

Although the divergence in the vertex can be written as a multiplication of the bare vertex by a divergent product of dressing functions, a variety of studies have shown that in general such a dressed vertex is not sufficient in its application to the Dyson-Schwinger equations to retrieve the correct divergence and the multiplicative renormalizability of the propagators. 

The DSE for the quark dressing function is:
\be\label{ZDSE}
\frac{1}{Z(p^2)} = 1 - \frac{\pi\alpha}{p^2} \intq{k} \Trace \left [ \gs{p} \; \Gamma_\mu(k,p) \gs{k} \, \gamma_\nu \right ] \, D^{\mu\nu}(q) \, \frac{Z(k^2)}{k^2} \,.
\ee

Expanding the RHS to order $\alpha$, i.e. replacing the full vertex by its bare analog, yields
\be\label{invZ}
\frac{1}{Z(p^2)} = 1 - \frac{\alpha\xi}{4\pi} \ln \frac{p^2}{\Lambda^2}\,.
\ee
To ${\cal O}(\alpha)$ this gives
\be\label{Zp}
Z(p^2) = 1 + \frac{\alpha\xi}{4\pi} \ln \frac{p^2}{\Lambda^2}\,,
\ee
and multiplicative renormalizability tells us that
\be\label{Zpow}
Z(p^2) = \left(\frac{p^2}{\Lambda^2} \right)^{\frac{\alpha\xi}{4\pi}} \,,
\ee
after resummation of the leading-logarithm contributions.

The question now is: how is the correct $\ln^2$-term produced such that the series \eqref{Zp} is continued in a way consistent with the MR solution \eqref{Zpow}?  This was discussed previously in ref.\ \cite{Curtis:1990zs}, and the mechanism is clarified in detail below.

For large momentum the leading-logarithm one-loop fermion-photon vertex is \cite{Curtis:1990zs}:
\be\label{Pvertex}
\Gamma_\mu(k,p) \stackrel{k^2\gg p^2}{=} \gamma_\mu \left ( 1- \frac{\alpha\xi}{4\pi}\ln\frac{k^2}{\Lambda^2} \right ) - \frac{\alpha}{4\pi} \bigg( \gs{p} \, \gamma_\mu \gs{k} + (\xi-1) k_\mu \gs{p} \bigg) \frac{1}{k^2} \ln\frac{k^2}{p^2} \,.
\ee
Note that the ultraviolet divergent part of the vertex ($\Lambda$-dependent) is exclusively contained in the term proportional to $\gamma_\mu$ of the vertex, and for large $k$ the vertex correction is given by $1/Z(k^2)$. Nevertheless, the divergent part of the vertex is {\it not} sufficient to generate the correct divergence of the propagator in the DSE, nor is it capable to produce an MR solution: the finite, momentum dependent part of the vertex plays a crucial role.

The next-to-leading order term of $Z(p^2)$ is generated by substituting the ${\cal O}(\alpha)$ dressing function \eqref{Zp} and vertex \eqref{Pvertex} (neglecting subleading parts independent of $\xi$) in Eq.\ \eqref{ZDSE}. After performing the angular integrals, the ultraviolet integral becomes 
\be\label{Z2l}
\frac{1}{Z(p^2)} = 1 + \frac{\alpha\xi}{4\pi} \int_{p^2}^{\Lambda^2} \frac{d k^2}{k^2} + \left(\frac{\alpha\xi}{4\pi}\right)^2 \int_{p^2}^{\Lambda^2} \frac{d k^2}{k^2} 
\Bigg[ \underbrace{\left( - \ln \frac{k^2}{\Lambda^2} + \ln \frac{k^2}{p^2} \right)}_{\Gamma^{(1)} Z^{(0)}}  +  \underbrace{\ln \frac{k^2}{\Lambda^2}}_{\Gamma^{(0)} Z^{(1)}} \Bigg] + \cdots \,,
\ee
where $\Gamma^{(0)}$, $Z^{(0)}$, $\Gamma^{(1)}$, $Z^{(1)}$ respectively correspond to the bare and one-loop corrections to the vertex and dressing function. Integration yields the dressing function to two-loop:
\be
\frac{1}{Z(p^2)} = 1 - \frac{\alpha\xi}{4\pi} \ln \frac{p^2}{\Lambda^2} + \frac{1}{2} \left( \frac{\alpha\xi}{4\pi} \ln \frac{p^2}{\Lambda^2}  \right)^2  + \cdots \,.
\ee
These are the first three terms of the Taylor expansion of the inverse of Eq.\ \eqref{Zpow}.
From Eq.\ \eqref{Z2l} it is clear that only taking into account the divergent parts of the one-loop vertex and dressing function is not sufficient to reproduce the correct two-loop contribution to the dressing function, which is crucial to ensure the resummability of the series required by multiplicative renormalizability. Therefore, DSE truncations where the full vertex is simply approximated by its divergent part, i.e. bare vertex multiplied with dressing functions, will not preserve the multiplicative renormalizability of the propagators. 

How can MR be preserved and the correct divergence be reproduced in a truncation using only the $\gamma_\mu$ part of the vertex?

The self-energy integral of Eq.\ \eqref{ZDSE} can be written as
\begin{align}
\intq{k} K(x,y,z) Z(y) &= \intq{k} [Z^2(y)  K(x,y,z)] \frac{1}{Z(y)} \notag\\
&\stackrel{\text{UV}}{=} \frac{\alpha\xi}{4\pi} \int_{x}^{\Lambda^2} \frac{d y}{y} \Bigg [  -1 - \frac{\alpha\xi}{4\pi} \underbrace{ \left(  \ln \frac{y}{\LUVsq} + \ln \frac{y}{x} \right) }_{(a)} \Bigg ]  \left( 1 - \frac{\alpha\xi}{4\pi} \ln \frac{y}{\LUVsq} \right) \,,
\end{align}
where $Z^2 K$ and $1/Z$ were expanded to ${\cal O}(\alpha)$.
The interesting observation is that the factor (a) yields a vanishing $\ln^2$ integral contribution. This explicitly shows that the $\ln^2$-correction to $1/Z(x)$ is correctly generated by the integral over the bare vertex kernel $K_0$ times the $\ln$-correction of $1/Z(y)$, i.e.
\be\label{truncation}
\intq{k} K(x,y,z) Z(y) = \intq{k}  K_0(x,y,z) \frac{1}{Z(y)} \qquad \text{to ${\cal O}(\ln^2)$} \,.
\ee

In fact, using the Taylor expansion of the power law \eqref{Zpow} it is easy to see that the RHS of Eq.\ \eqref{truncation} reproduces term by term the leading-logarithm series for $1/Z(x)$,  and Eq.\ \eqref{truncation}  is valid at each order in perturbation theory. This means that the MR truncation correctly reproduces the resummed leading-logarithm results.
Note that this does not imply that the kernels $Z^2 K$ and $K_0$ are identical, but just that they yield the same ultraviolet integral contributions. 
In the same way, the MR truncation can be shown to reproduce the resummed leading-logarithm results of the gluon and ghost propagators in QCD.

The MR truncation of Sect.\ \ref{Sect:Truncation} assumes that the use of Eq.\ \eqref{truncation} can reasonably be extended to all external momenta, at least in a slightly more general form, as it ensures that multiplicative renormalizability is satisfied over the complete momentum domain.

\section{Feynman Integration Trick}

To solve the infrared integrals of Sect.\ \ref{Sect:IR} using the integrals given in \ref{App:tensorintegrals}, it is helpful to transform denominator products of squared momenta in the kernels using the following equality:
\be\label{FT1}
\frac{1}{{p^2}^{a} {q^2}^{b}} = \frac{\Gamma(a+b)}{\Gamma(a)\Gamma(b)} \int_0^1 dt \;  t^{a-1} (1-t)^{b-1} \frac{1}{\left[ t p^2 + (1-t) q^2 \right]^{a+b}} \,,
\ee
which can be generalized as
\be\label{FT2}
\frac{1}{{q_1^2}^{m_1} {q_2^2}^{m_2} \ldots {q_n^2}^{m_n}} = \frac{\Gamma(m_1+m_2+\cdots+m_n)}{\Gamma(m_1)\Gamma(m_2)\ldots\Gamma(m_n)}  \int_0^1 dt_1 dt_2 \ldots dt_n \;\delta\left(\sum t_i-1\right) \; \frac{ \prod t_i^{m_i-1}}{\left[ \sum t_i q_i^2 \right]^{\sum m_i}} \,.
\ee

Eventually the Feynman parameter $t$ is integrated out using the well-known formula:
\be\label{integt}
\int_0^1 dt \; t^a (1-t)^b = \frac{\Gamma(a+1)\Gamma(b+1)}{\Gamma(a+b+2)} \,.
\ee

\section{Integrals}
\label{App:tensorintegrals}

The tensor integrals needed to compute the infrared contributions of the one- and two-loop diagrams are \cite{Peskin:1995ev}
\begin{align}
\label{J0} J(n,\Delta) &\equiv \intq{\ell} \; \frac{1}{(\ell^2+\Delta)^n} =  \frac{1}{16\pi^2}\frac{\Gamma(n-2)}{\Gamma(n)}\Delta^{2-n} \,, \\
\label{J2} J_{\mu\nu}(n,\Delta) &\equiv \intq{\ell} \; \frac{\ell_\mu \ell_\nu }{(\ell^2+\Delta)^n} =  \frac{1}{16\pi^2}\frac{\Gamma(n-3)}{\Gamma(n)} \Delta^{3-n}\frac{\delta_{\mu\nu}}{2} \,, \\
\label{J4} J_{\mu\nu\rho\sigma}(n,\Delta) &\equiv \intq{\ell} \;  \frac{\ell_\mu \ell_\nu \ell_\rho \ell_\sigma}{(\ell^2+\Delta)^n} = \frac{1}{16\pi^2}\frac{\Gamma(n-4)}{\Gamma(n)}\Delta^{4-n}\frac{1}{4}(\delta_{\mu\nu}\delta_{\rho\sigma}+\delta_{\mu\rho}\delta_{\nu\sigma}+\delta_{\mu\sigma}\delta_{\nu\rho}) \,, \\
\label{J6} 
\begin{split}
J_{\mu\nu\rho\sigma\tau\gamma}(n,\Delta) &\equiv \intq{\ell}  \;  \frac{\ell_\mu \ell_\nu \ell_\rho \ell_\sigma \ell_\tau \ell_\gamma}{(\ell^2+\Delta)^n} = \frac{1}{16\pi^2}\frac{\Gamma(n-5)}{\Gamma(n)}\Delta^{5-n} 
\frac{1}{8} \big[(\delta_{\mu\nu}\delta_{\rho\sigma}  + \delta_{\mu\rho}\delta_{\nu\sigma} + \delta_{\mu\sigma}\delta_{\nu\rho})\delta_{\tau\gamma}
\\
&  + (\delta_{\gamma\nu}\delta_{\rho\sigma} + \delta_{\gamma\rho}\delta_{\nu\sigma} + \delta_{\gamma\sigma}\delta_{\nu\rho})\delta_{\tau\mu} 
+ (\delta_{\mu\gamma}\delta_{\rho\sigma} + \delta_{\mu\rho}\delta_{\gamma\sigma} + \delta_{\mu\sigma}\delta_{\gamma\rho})\delta_{\tau\nu} \\
& + (\delta_{\mu\nu}\delta_{\gamma\sigma} + \delta_{\mu\gamma}\delta_{\nu\sigma} + \delta_{\mu\sigma}\delta_{\nu\gamma})\delta_{\tau\rho} 
+ (\delta_{\mu\nu}\delta_{\rho\gamma} + \delta_{\mu\rho}\delta_{\nu\gamma} + \delta_{\mu\gamma}\delta_{\nu\rho})\delta_{\tau\sigma} \big] \,, 
\end{split}\raisetag{20pt}
\end{align}
and the odd-rank tensor integrals vanish because of symmetry properties.

By applying the Feynman integration trick \eqref{FT1} and using the integration formula (\ref{J0}), a typical scalar integral, often needed in one-loop integrals, is given by
\be\label{Iab}
I(a,b) = \int \frac{d^4 q}{(2\pi)^4}\, \frac{1}{y^{a} z^{b}} 
= \frac{1}{16\pi^2} \frac{\Gamma(2-a)\Gamma(2-b)\Gamma(a+b-2)}{\Gamma(a)\Gamma(b)\Gamma(4-a-b)} \, x^{2-a-b}\,,
\ee
with $x=p^2$, $y=q^2$ and $z=(p-q)^2$.

\section{Effective Two-loop Kernels}
\label{App:Eff2loop}

As discussed in Sect.\ \ref{Sect:EffLoop} the two-loop diagrams are approximated by effective one-loop integrals in the numerical calculation. The kernel of the effective squint diagram can be written as
\be\label{Ksquint}
Q_4^{\text{eff}}(x,y,z) = {\cal A}(k2,k3) \left(\frac{\frac{-17}{y} - \frac{3\,y}{z^2} - \frac{4}{z}}{x} - 
  \frac{6}{z^2} + \frac{5\,x}{y\,z^2} - \frac{8}{y\,z} + 
  \frac{-36 + \frac{4\,y^2}{z^2} + \frac{12\,y}{z} + 
     \frac{20\,z}{y}}{x^2}\right) \,,
\ee
with 
\be
{\cal A}(k2,k3) = \frac{- \Gamma(1 + {k_2})\,\Gamma(1 + {k_3})\,\Gamma(-{k_2} - {k_3})\,
      \left( -5 + 3\,{{k_2}}^2  + 3\,{{k_3}}^2 \, + {k_2}\,{k_3} + {k_2}   + 
        {k_3} \right) }
    {256\,{\pi }^2\,\Gamma(2 - {k_2})\,
    \Gamma(2 - {k_3})\,
    \Gamma(3 + {k_2} + {k_3})} \,.
\ee
   
Kernel \eqref{Ksquint} contains potential quartic ($1/x^2$-term) and quadratic ($1/x$-term) divergences. Although these terms formally cancel after integration, their numerical implementation can yield serious problems because incomplete cancellations will give rise to residual unrenormalizable divergences \cite{Bloch:1995dd, Bloch:1995if}. A simple, elegant solution consists in writing the expression over a common denominator and then express $z$ as $x+y-2\sqrt{xy}\cos\theta$:
\be
\begin{split}
Q_4^{\text{eff}}(x,y,z) = {\cal A}(k2,k3)\;  \frac{1}{x\,y\,z^2} \; 
& [   24 y^2 \left( -1 + 4\cos^2\theta \right) - 20 y \sqrt{x y} \cos\theta \left( 1 + 8\cos^2\theta\right) \\
& + 4 x y \left( -7 + 43\cos^2\theta \right)
- 36\,x  {\sqrt{x\,y}}\,{\cos\theta }\,	
        ] \,.
\end{split}
\ee  
Once written in this form the simple trigonometric form of the remaining $1/x$-term ensures that the quadratic divergence will be cancelled properly. An analogous expression can be written for the sunset diagram.

\section{Numerical Method}
\label{App:NumMethod}

The numerical method employed to solve the set of coupled integral equations \eqref{subtracted} was first devised for the study of dynamical chiral symmetry breaking in strong coupling QED \cite{Bloch:1995dd}. 
After changing to spherical coordinates and integrating out two angles, the integral equations can be written as (dropping the subscript $R$ on the renormalized dressing functions)
\be
\label{spherical}
\begin{aligned}
\frac{1}{F(x)} &= \frac{1}{F_\sigma} + \int_0^\infty d y \, \int_0^\pi d\theta \,\, K_F[G,F](x,y,\theta) \,, \\
\frac{1}{G(x)} &= \frac{1}{G_\sigma} + \int_0^\infty d y \, \int_0^\pi d\theta \,\, K_G[G,F](x,y,\theta) \,.
\end{aligned}
\ee

The unknown dressing functions are approximated using Chebyshev expansions \cite{Numrec} in $\log(x)$ for $x \in [\epsilon^2,\Lambda^2]$:
\be\label{Cheb}
\begin{aligned}
\log F(x) &:= \frac{a_0}{2} + \sum_{j=1}^{N-1} a_j \, T_j(s(x)) \,, \\
\log G(x) &:= \frac{b_0}{2} + \sum_{j=1}^{N-1} b_j \, T_j(s(x)) \,,
\end{aligned}
\ee
where $\epsilon^2$ and $\Lambda^2$ are infrared and ultraviolet parameters introduced for numerical purposes only, and
\be
s(x) \equiv \frac{\log(x/\Lambda\epsilon)}{\log(\Lambda/\epsilon)} \,,
\ee
such that $s \in [-1,+1]$ when $x \in [\epsilon^2,\Lambda^2]$.
The number of Chebyshev polynomials is chosen between 30 and 60, depending on the steepness of the running coupling in the intermediate momentum regime and the required accuracy. In the infrared, for $x < \epsilon^2$, the dressing functions are taken to obey their asymptotic power law behaviors
\be\label{powlaw2}
\begin{aligned}
F(x) &:= A \, x^{2\kappa} \,,\\
G(x) &:= B \, x^{-\kappa} \,,
\end{aligned}
\ee
and in the ultraviolet, for $x > \Lambda^2$, the functions are frozen to their values at $\Lambda^2$ given by the Chebyshev expansions \eqref{Cheb}.

The system of equations \eqref{spherical} is required to be satisfied at $N$ external momenta
\be
x_i = \Lambda\epsilon\left(\frac{\Lambda}{\epsilon}\right)^{s_i} \quad , \quad
s_i = \cos\left[\frac{(i-1/2)\pi}{N}\right] \quad , \quad i = 1,\ldots,N \,,
\ee
where the $s_i$ are chosen to be the zeroes of the Chebyshev polynomial $T_N$ of degree $N$, and the integrals are approximated by suitable numerical quadratures:
\be
\label{discrete}
\begin{aligned}
\frac{1}{F(x_i)} &= \frac{1}{F_\sigma} + \sum_{j=1}^{N_y} \sum_{k=1}^{N_\theta} w_j \, w_k \, K_F[F,G](x_i,y_j,\theta_k) \,, \\
\frac{1}{G(x_i)} &= \frac{1}{G_\sigma} + \sum_{j=1}^{N_y} \sum_{k=1}^{N_\theta} w_j \, w_k \, K_G[F,G](x_i,y_j,\theta_k) \,,
\end{aligned}
\ee
with radial integration points $y_j$, $j=1,\ldots, N_y$, and angular integration points $\theta_k$, $k=1,\ldots,N_\theta$. Note that the nodes $y_j$ can depend on the external momentum $x_i$.

The radial integrals are split into three parts: $[0, \epsilon^2]$, $[\epsilon^2, \Lambda^2]$, and $[\Lambda^2,\infty]$. An accurate evaluation of the infrared integrals $[0,\epsilon^2]$ is important to reproduce the infrared power laws, and they are solved using a Gauss-Legendre integration with 64 points.
The major integration range $[\epsilon^2, \Lambda^2]$ has to be handled carefully as the integrand is not smooth at $y=x$. A composite Gauss-Legendre rule is used (after transformation to $t=\log(y)$), with main nodes located at $\epsilon^2$, $\Lambda^2$ and the $N$ values of external momenta, and an $M$-point Gauss-Legendre rule is used between adjacent nodes. The integrand is concentrated around $y=x$, and hence the number of points in the elementary rule is adapted accordingly: a high accuracy rule with $M=32$ is used when the integration frame overlaps with the interval $[x/10,10x]$, while otherwise $M=4$. For the subtracted equations the ultraviolet integral $[\Lambda^2,\infty]$ is negligible. The angular integrals are computed using a $N_\theta$-point Gauss-Legendre integration, with $N_\theta=33$.

The method described above transforms the set of coupled integral equations \eqref{spherical}, into a set of $2 N$ non-linear algebraic equations \eqref{discrete} for the Chebyshev coefficients $a_j, b_j$ of the dressing functions. Symbolically this can be written as
\be
\left\{
\begin{aligned}
f(x)[F,G] &= 0  \\
g(x)[F,G] &= 0
\end{aligned} \right.
\qquad\Longrightarrow\qquad
\left\{
\begin{aligned}
\tilde f_i[a_j,b_j] &= 0 \\
\tilde g_i[a_j,b_j] &= 0 
\end{aligned} \right. \quad .
\ee

The latter set of equations is solved using the Newton iterative method, which is quadratically convergent for starting guesses close enough to the solutions. Successive iterations are given by
\be\label{Newton}
\left\{
\begin{aligned}
a_j^{n+1} &= a_j^n - \Delta a_j^{n+1} \\
b_j^{n+1} &= b_j^n - \Delta b_j^{n+1} 
\end{aligned}
\right. \quad ,
\ee
where the increments $\Delta a_j$ and $\Delta b_j$ are solutions of the set of $2N$ linear equations
\be
\left\{
\begin{aligned}
\frac{\partial \tilde f_i[a_j^n,b_j^n]}{\partial a_j}\,\Delta a_j^{n+1} 
+ \frac{\partial \tilde f_i[a_j^n,b_j^n]}{\partial b_j}\,\Delta b_j^{n+1} &= \tilde f_i[a_j^n,b_j^n] \\ 
\frac{\partial \tilde g_i[a_j^n,b_j^n]}{\partial a_j}\,\Delta a_j^{n+1} 
+ \frac{\partial \tilde g_i[a_j^n,b_j^n]}{\partial b_j}\,\Delta b_j^{n+1} &= \tilde g_i[a_j^n,b_j^n] \,.
\end{aligned} , \quad i=1\ldots N,
\right.
\ee
and summation over the repeated index $j$ is assumed.
The quadratic convergence property is clearly observed in the numerical implementation as the distance between successive iterations decreases by an order of magnitude at each iteration step in the vicinity of the solution. Hence, the method is fast and gives a reliable estimate of the accuracy of the solutions. 

Note that the quadratic convergence only sets in "close" enough to the exact solution, and good initial guesses dramatically speed up the convergence rate. These can easily be constructed, as the infrared and ultraviolet asymptotic behaviors of the dressing functions are known. Typical starting guesses are
\be\label{guesses}
\begin{aligned}
\tilde F(x) &= A \left(\frac{x}{x/x_t+1}\right)^{2\kappa} \left(\frac{\tilde \alpha(x/x_t)}{\alpha_0}\right)^{13/22} \,, \\
\tilde G(x) &= B \left(\frac{x}{x/x_t+1}\right)^{-\kappa} \left(\frac{\tilde \alpha(x/x_t)}{\alpha_0}\right)^{9/44} \,,
\end{aligned}
\ee
with
\[
\tilde \alpha(t) = \frac{1}{c_0+t^2} \left[ c_0 \, \alpha_0  + \frac{4\pi}{\beta_0}\left(\frac{1}{\log t}+\frac{1}{1-t}\right) t^2 \right] \,,
\]
where $c_0$ and $x_t$ can be tuned to improve the starting guess. 
The procedure \eqref{Newton} is iterated till the convergence criterion $\max_j(|\Delta a_j^{n+1}|) < 10^{-5} \max_j(|a_j^{n+1}|)$ and $\max_j(|\Delta b_j^{n+1}|) < 10^{-5} \max_j(|b_j^{n+1}|)$ is satisfied.
Starting from pure power laws \eqref{powlaw2} for all momenta typically requires about 20 iterations to converge, while the more refined starting guesses \eqref{guesses} reduce the iteration procedure to 4 iteration steps.

The total accuracy of the method is given by a combination of the accuracies of the Chebyshev expansion, the quadrature rules, and the Newton iteration method.

\bibliography{MRfull}

\end{document}